\RequirePackage[l2tabu]{nag}
\documentclass{article}

\usepackage[utf8]{inputenc}
\usepackage[T1]{fontenc}

\usepackage{geometry}
\usepackage{color}
\usepackage{textcomp}
\usepackage{float}
\usepackage{graphicx}
\usepackage{caption}
\usepackage{subcaption}
\usepackage{amsmath}
\usepackage{amssymb}
\usepackage{color}
\usepackage{placeins}
\usepackage{algorithm}
\usepackage{algpseudocode}
\usepackage{dcolumn}%
\usepackage{bm}%

\usepackage[round]{natbib}
\usepackage[sectionbib]{bibunits}
\defaultbibliographystyle{plainnat}
\usepackage{enumitem}

\usepackage{graphicx}
\usepackage{amsmath}
\usepackage[version=4]{mhchem}
\usepackage{siunitx}
\usepackage{longtable,tabularx}
\usepackage{lmodern}
\usepackage[tracking=true,kerning=true,final]{microtype}

\usepackage{hyperref}
\hypersetup{colorlinks=true, linkcolor=blue,  anchorcolor=blue, citecolor=blue, filecolor=blue, menucolor=blue, urlcolor=blue}
\hypersetup{pdfauthor={},pdftitle={}}

\usepackage{caption}   
\usepackage{subcaption}
\usepackage{color}
\usepackage{graphicx}
\graphicspath{{./}{./figs/}}

\usepackage{amsfonts}
\usepackage{bm}
\usepackage{mathtools}

\title{Harnessing the instability mechanisms in airfoil flow for the data-driven forecasting of extreme events}

\author{Benedikt Barthel, Themistoklis Sapsis
\thanks{Corresponding author: \href{mailto:bbarthel@mit.edu}{bbarthel@mit.edu}
}\\
Department of Mechanical Engineering,
\\ Massachusetts Institute of Technology, \\
77 Massachusetts Ave., Cambridge, MA 02139}
\date{\today}

\usepackage{amsmath}
\usepackage{amssymb}

\usepackage[normalem]{ulem}

\begin{document}

\maketitle\ 
\begin{abstract}
This work addresses the data-driven forecasting of extreme events in the flow over a static airfoil. For certain Reynolds numbers and flow configurations, airfoils are subject to sporadic high amplitude fluctuations in the aerodynamic forces. These extreme excursions may be seen as prototypical examples of the kind of unsteady and intermittent dynamics relevant to the flow around airfoils and wings in a variety of laboratory and real-world applications. Here we investigate the instability mechanisms at the heart of these extreme events, and how knowledge thereof may be harnessed for efficient data driven forecasting. Through a wavelet and spectral analysis of the flow we find that the extreme events arise due to the instability of a specific frequency component distinct from the vortex shedding mode. During these events this extreme event manifold draws energy from the energetically dominant vortex shedding flow and undergoes an abrupt inverse cascade of energy transfer from small to large scales. We also investigate the spatial dependence of the temporal correlation and mutual information between the surface pressure and the aerodynamic forces, with the aim of identifying regions of the airfoil amenable to sparse sensing and the efficient forecasting of extremes. Building on previous work on predictive machine learning models, we show that relying solely on the mutual information for optimal sensor placement fails to improve model prediction over uniform or random sensor placement. However, we show that by isolating the extreme event frequency component offline through a wavelet transform we are able to circumvent the requirement for a recursive long-short term memory (LSTM) network -- resulting in a significant reduction in computational complexity over the previous state of the art. Using the wavelet pre-processed data in conjunction with an extreme event-tailored loss function we find that our model is capable of forecasting extreme events using only three pressure sensors. Furthermore, we find our model to be robust to sensor location -- showing promise for the use of our model in dynamically varying applications. 
\end{abstract}

\section{Introduction}
Many engineering systems are subject to rare high-amplitude fluctuations commonly referred to as \textit{extreme events} \citep{sapsis_statistics_2021}. While here we focus primarily on fluid-structure interactions, such events occur in a wide variety of systems ranging from climate systems to stock markets. Although rare, events like gusts or rogue waves have a disproportionate affect on the fatigue life of aircraft, naval vessels, or marine infrastructure. Due to their rare nature, the prediction of these events is inherently challenging, especially as they often occur in complex systems were the physical mechanisms are unknown \citep{Farazmand2019}. Various authors have ventured to address this problem through strategies such as optimal sampling \citep{mohamad_sequential_2018,sapsis_output-weighted_2020, blanchard_output-weighted_2021} or training strategies which preferentially amplify rare events \citep{guth_machine_2019,qi_using_2020,rudy_prediction_2022}. 

Two dimensional airfoil flow is one of the canonical test cases for the dynamics of fluid-structure interaction and has been the subject of extensive study for decades \citep{lissaman_low-reynolds-number_1983,kim_low-reynolds-number_2012}. However, the advent of machine learning and data driven techniques has unlocked new lenses to study this and other classical problems in the field of fluid dynamics \citep{brenner_perspective_2019,fukami_assessment_2020,brunton_machine_2020}. In particular, several authors including \citet{gomez_unsteady_2019, maulik_probabilistic_2020,rudy_prediction_2022} have proposed neural network models for the reconstruction of the flow from sparse measurements. Data driven prediction's based the airfoil surface pressure have been of particular interest due to it's practical measurability. Most aerospace applications will require predictions made from sparse arrays of sensors, and thus effective strategies for the optimal distribution of pressure sensors is of critical importance. \citet{hou_machine-learning-based_2019,le_provost_deep_2020} used both data assimilation and convolutional and recursive neural networks for the context of prediction of the leading edge suction parameter (LESP), while \citet{rudy_prediction_2022} investigated a range of neural network models for the prediction of the drag coefficient. 

Oscillator flows such as airfoil flows are generally insensitive to noise and exhibit multiple characteristic time scales \citep{williamson_vortex_1996,symon_non-normality_2018}. This makes them an ideal candidate for the study of multi-scale slow-fast type extreme events -- as classified by \citet{Farazmand19}.
From this perspective, the Reynolds number regime $\mathcal{O} \sim \left(10^4\right)$ is of particular interest. This regime lies between steady laminar and fully turbulent regimes, and is especially susceptible to highly nontrivial dynamics which depend significantly on angle of attack and Reynolds number \citep{wang_turbulent_2014,gopalakrishnan_meena_airfoil-wake_2018,menon_aerodynamic_2020}. One tool for the study of such systems is the continuous wavelet transform, which in a manner analogous to the short time Fourier transform, quantifies the time varying strength of a signal's frequency components  \citep{addison_illustrated_2016,mojahed_new_2021}. Wavelet analysis has previously been used in the context of extreme events by \citet{cousins_sapsis,cousins_reduced-order_2016,bayindir} and \citet{wavelet_sri} for the detection of rogue waves and turbulent bursts in pipe flow respectively. The majority of these studies have focused on spatial wavelet transforms, while here we perform a temporal analysis. Additionally, the combination of wavelet analysis with machine learning models using the type of output weighted strategies discussed above remains largely unexplored. 

This work conducts an investigation of the physical mechanisms driving the extreme bursting events observed in the flow over a two-dimensional airfoil at $R = 17,500$ at constant angle of attack.  We build on previous work by \citet{rudy_prediction_2022} who studied the data driven reconstruction of this flow using a range of neural network architectures. We exploit the findings of our analysis to design extreme event tailored -- also referred to as output-weighted \citep{sapsis_output-weighted_2020} -- data processing and training strategies for the efficient data-driven prediction of extreme events from optimal sparse sampling of the surface pressure. For a broad discussion of output-weighted strategies for neural networks applied to this and other systems see \citet{rudy_output-weighted_2021}.

The rest of the paper is organized as follows. In \S\ref{sec:problem} we describe the problem under investigation. In \S\ref{sec:stats} we perform a statistical analysis of the data and describe the physical mechanisms driving the extreme events, and how these manifest in the observed data. In \S\ref{sec:offline} we discuss the limitations of offline optimal sensing algorithms. The main results of this work are then presented in \S\ref{sec:wavelet}, and we provide some discussion of our findings in \S\ref{sec:discussion}.



\section{Problem Description}\label{sec:problem}
We consider a 2D direct numerical simulation of an incompressible flow around a NACA 4412 airfoil at an angle of attack $\alpha = 5^{\circ}$ and a cord length based Reynolds number $R = 17,500$. The flow is governed by the Navier-Stokes and continuity equations,
\begin{equation}
    \frac{\partial \mathbf{u}}{\partial t} +\mathbf{u}\cdot\nabla\mathbf{u}-\frac{1}{R}\nabla^2\mathbf{u} + \nabla p =0,
\end{equation}
\begin{equation}
    \nabla\cdot\mathbf{u} =0,
\end{equation}
where $\mathbf{u} \equiv [u(x,y,t),v(x,y,t)]$ is the velocity, $p(x,y,t)$ is the pressure field, $t$ is time, and $x,y$ are the spatial dimensions parallel and perpendicular to the free stream respectively. The simulation is carried out using the open source spectral element code Nek5000 developed by \citet{fischer_nek5000_2008} with 4368 elements at spectral order 7 and a convective outflow boundary condition \citep{dong_convective-like_2015}. We use the same data set as \citet{rudy_prediction_2022} who report that further refinement of the numerical grid did not meaningfully alter the results. At this Reynolds number the flow is susceptible to intermittent, yet non-periodic turbulent bursts which manifest as high amplitude fluctuations in the drag coefficient. Therefore we focus on two observables: the surface pressure and the drag coefficient. The former is a practically measurable quantity and will serve as the input to our model, while the latter encodes the extreme events and will serve as the model output. 

\begin{figure}
    \centering
    \includegraphics[trim = 20 0 0 0,scale = 0.52]{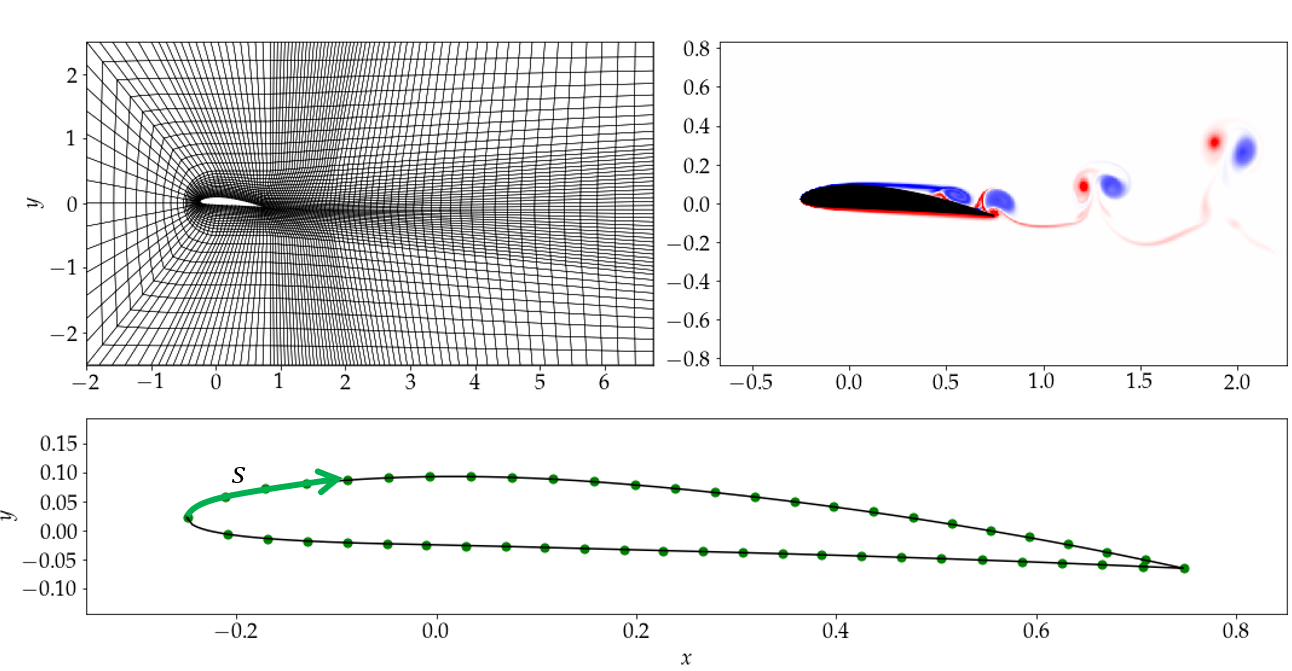}
    \caption{From top left: computational grid, snapshot of vorticity and airfoil geometry, with arclength measure $s$. Dots represent every other sensor location. Image adapted with permission from \citet{rudy_prediction_2022}.}
    \label{fig:airfoil}
\end{figure}

Throughout this work we define $s$ as a generalized measure of arc length measured clockwise from the leading edge (as shown in figure \ref{fig:airfoil}). For example,  $s\in[0,0.5)$ refers to the upper surface of the airfoil and $s\in[0.5,1)$ refers to under side. The surface pressure is saved at 100 equally spaced locations around the airfoil surface. A visualization of the airfoil flow, the simulation grid and the arc length measure are summarized in figure \ref{fig:airfoil}, and we refer the interested reader to \citet{rudy_prediction_2022} for a more detailed discussion of the numerical method.

\begin{figure}
    \centering
    \includegraphics[trim =  180 0 0 0, scale =0.18]{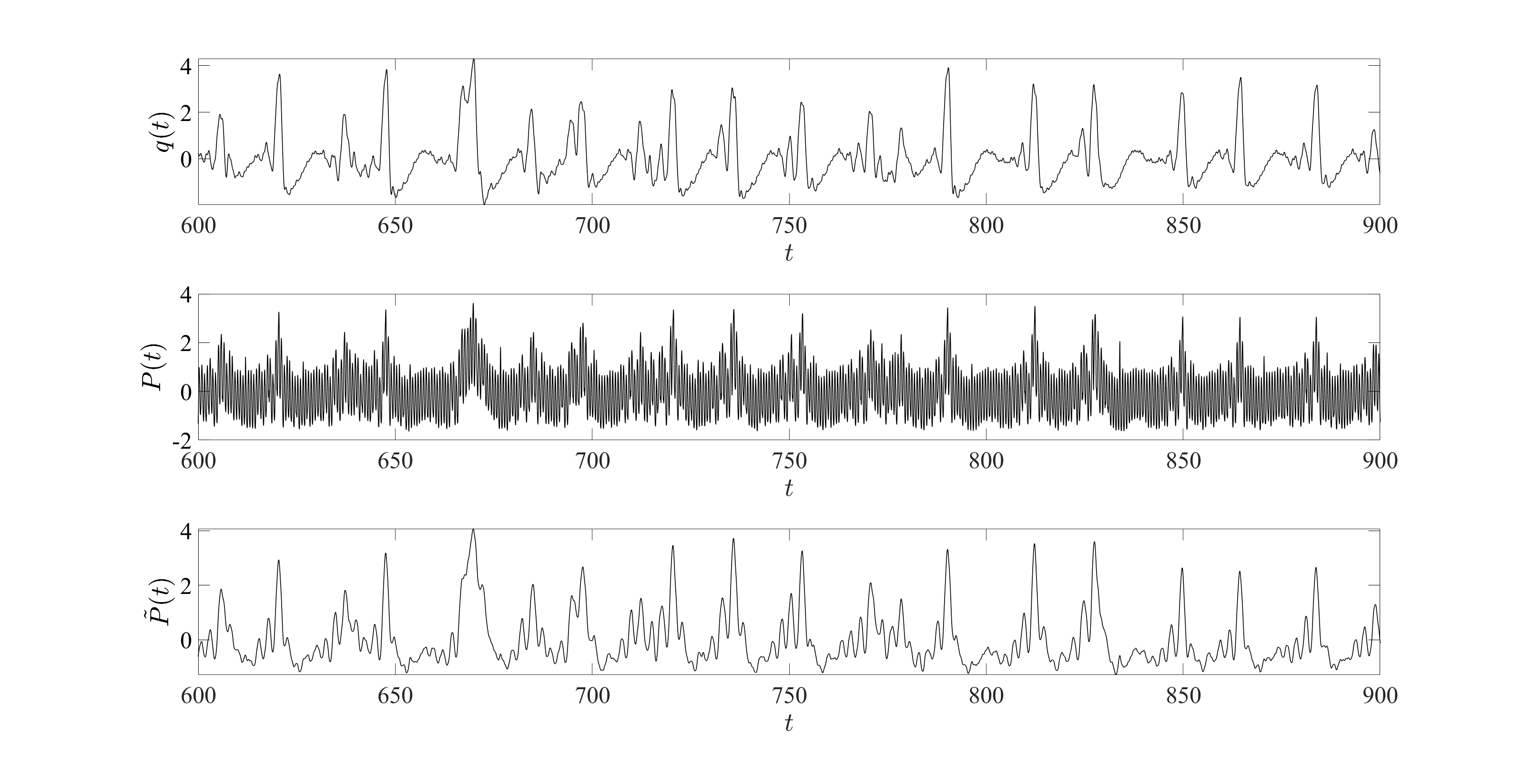}
    \caption{Illustrative example of the data. From top to bottom: filtered drag coefficient, raw pressure signal, and filtered pressure signal. Pressure data is taken at a single representative sensor location $25\%$ of the way along the upper surface. }
    \label{fig:example_data}
\end{figure}

The aerodynamic forces are computed using skin friction and surface pressure according to
\begin{equation}
    \mathbf{F}(t) = \oint_s \left(\boldsymbol{\tau}(s,t) + \mathbf{n}P(s,t)\right) ds = L(t)\hat{\mathbf{e}}_x + D(t)\hat{\mathbf{e}}_y.
\end{equation}
Here $t$ is time, and $x$ and $y$ represent the directions parallel and normal to the free stream respectively. The lift and drag coefficients are then defined as 
\begin{equation}
    C_L(t) \equiv \frac{L(t)}{\rho U_{\infty}^2c},  ~\ ~\ ~\ ~\ ~\ ~\ ~\ ~\ ~\ C_D(t) \equiv \frac{D(t)}{\rho U_{\infty}^2c}.
\end{equation}
To distinguish the extreme events from the background vortex shedding we apply a Gaussian smoothing operation to the time series of the drag coefficient to extract the non-periodic behaviour,
\begin{equation}
    q(t) \equiv \left(K*C_D \right)(t).
\end{equation}
where $K(t') \propto exp\left(-(t'/2 f_{v})^2 \right)$ is a Gaussian smoothing kernel. $f_{v} = 1.44$ is the most energetic frequency and corresponds to periodic vortex shedding. Moving forward we simply refer to $q(t)$ as the drag. In addition to the raw pressure signal, we also consider a version of the pressure with the same Gaussian filter applied,
\begin{equation}
\Tilde{P}(s,t) \equiv \left(K*P(s,t) \right)(s,t).
\end{equation}
We refer to $P(s,t)$ and $\tilde{P}(s,t)$ as the raw and filtered pressures respectively. In general we will consider the pressure measured at subset of discrete sensor locations, and thus treat the surface pressure as a vector valued quantity $\mathbf{P}(t) \in \mathbb{R}^{n}$, where $n$ is the number of sensors. An illustrative example of the drag as well as the raw and filtered pressures is shown in figure \ref{fig:example_data}.

This flow was previously investigated by \citet{rudy_prediction_2022} using a deep long-short term memory (LSTM) network. Those authors considered a variety of input observables, and found that the extreme events could be predicted from a range of different observables including full and reduced order descriptions of the flow field as well as surface pressure. This suggest that the extreme events are a result of an underlying physical instability inherent in the governing equations. The primary focus of this work is to identify and exploit this mechanism for optimal sensing and forecasting. Practically, we aim to predict future extreme events observed in the drag from sparse measurements of the surface pressure as efficiently as possible. In other words we seek a data driven map
\begin{equation}\label{map}
    \mathbf{P}(t) \rightarrow q(t+\tau)
\end{equation}
for maximum lead time $\tau$, with minimal $ \operatorname{dim}\left(\mathbf{P}\right)$, and at minimal computational cost.


\section{Statistical Analysis and Physical Mechanisms of Extremes}\label{sec:stats}
In order to gain a deeper insight into the dynamics of the flow and mechanisms driving the extreme events, we first perform a detailed statistical analysis of the data. We first analyze the surface pressure, as this will serve as the basis of our modeling efforts, then in \S\ref{sec:flowfield} we analyze the vorticity field -- both globally and locally along the airfoil surface -- to further probe the extreme event dynamics.
Before presenting our results, we first review some definitions we use throughout the following sections. 
For a signal $x(t)$ with discrete values $x_i$ and distribution $X$, we define the mean $\mu_x$, variance $\sigma^2_x$, and the probability density function $f_X(x)$. For two signals $x(t)$ and $y(t)$ the covariance is defined as
\begin{equation}
    \sigma_{xy} \equiv \operatorname{cov}(X,Y) = \frac{1}{n-1} \sum_{i=1}^n \left(x_i - \mu_x\right) \left(y_i - \mu_y\right).
\end{equation}
To further quantify the connection between two signals we also make use the mutual information (MI) defined as
\begin{equation}\label{MI}
    I(X,Y) \equiv \int_y \int_x f_{X,Y}(x,y)\log\left(\frac{f_{X,Y}(x,y)}{f_X(x)f_Y(y)}\right) dx dy,
\end{equation}
where $f_{X,Y}$ is the joint probability density function of $X$ and $Y$. The MI is the Kullback–Leibler (KL) divergence between the joint probability distribution and the product of the marginal probability distributions -- it quantifies the error in the assumption that two distributions $X$, and $Y$ are uncorrelated.
We also propose the ``extreme event conditioned mutual information'', defined as the mutual information integrated only over values of the output greater than two standard deviations from the mean: $y>2\sigma_y$ -- all values of the input $X$ are included --
\begin{equation}
    I_{EE}(X,Y) \equiv \int_x \int_{y>2\sigma_y} f_{X,Y}(x,y)\log\left(\frac{f_{X,Y}(x,y)}{f_X(x)p_Y(y)} \right) dx dy.
\end{equation}\label{MI_EE}
We choose a cut-off of two standard deviations, however we found that the results were not sensitive to changes of $\pm\sigma$. For the results presented in this section the probability density functions in (\ref{MI}) and (\ref{MI_EE}) are approximated using Monte Carlo estimation using 50,000 samples and the relevant integrals are then carried out using trapezoidal integration.

\subsection{Mutual Information Structure}\label{sec:MIstruct}
\begin{figure}
    \centering
    \subfloat[]{%
    \includegraphics[trim = 90 0 0 0, scale = 0.32]{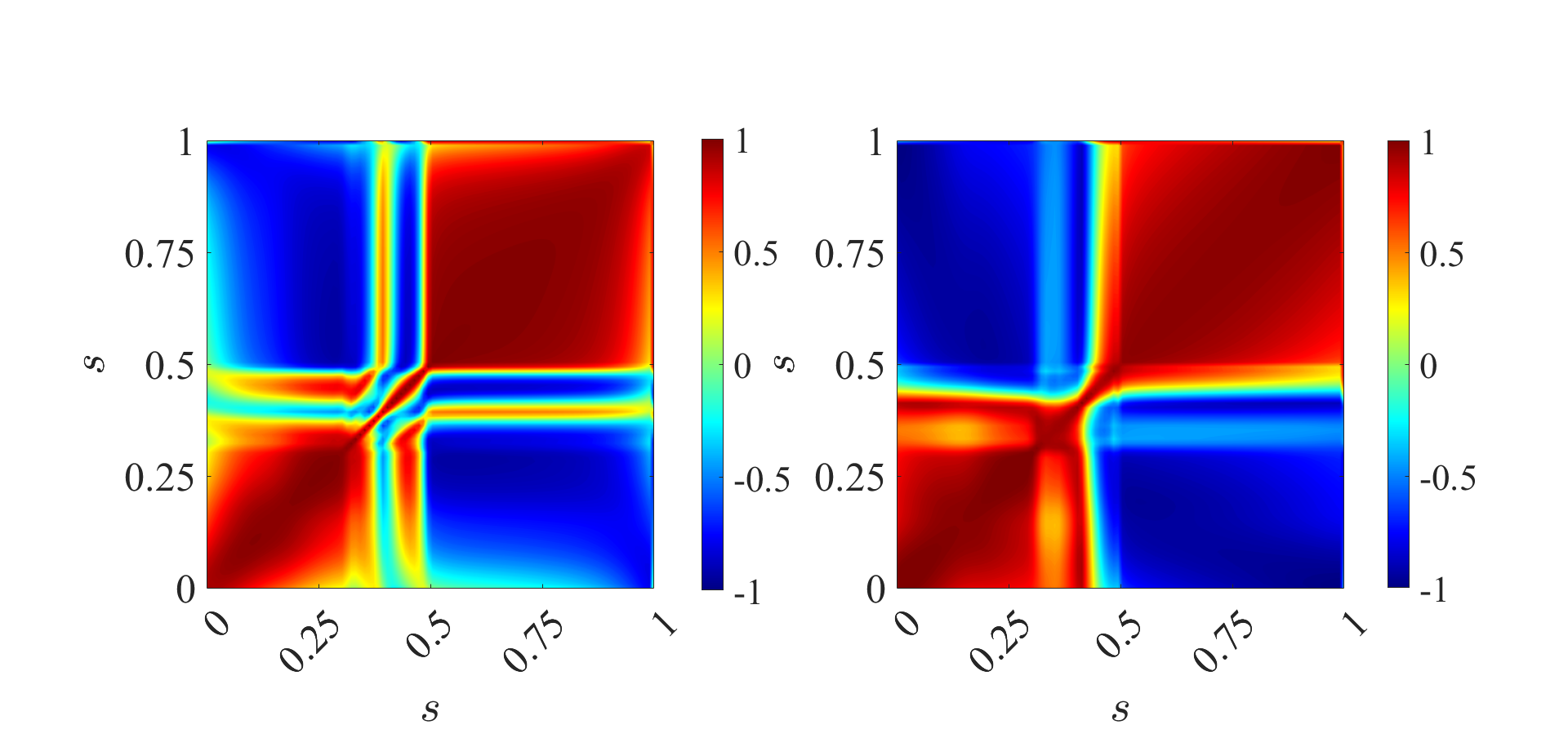}%
    \label{fig:P_COV_map}
}

\subfloat[]{%
    \includegraphics[trim = 90 0 0 0, scale = 0.32]{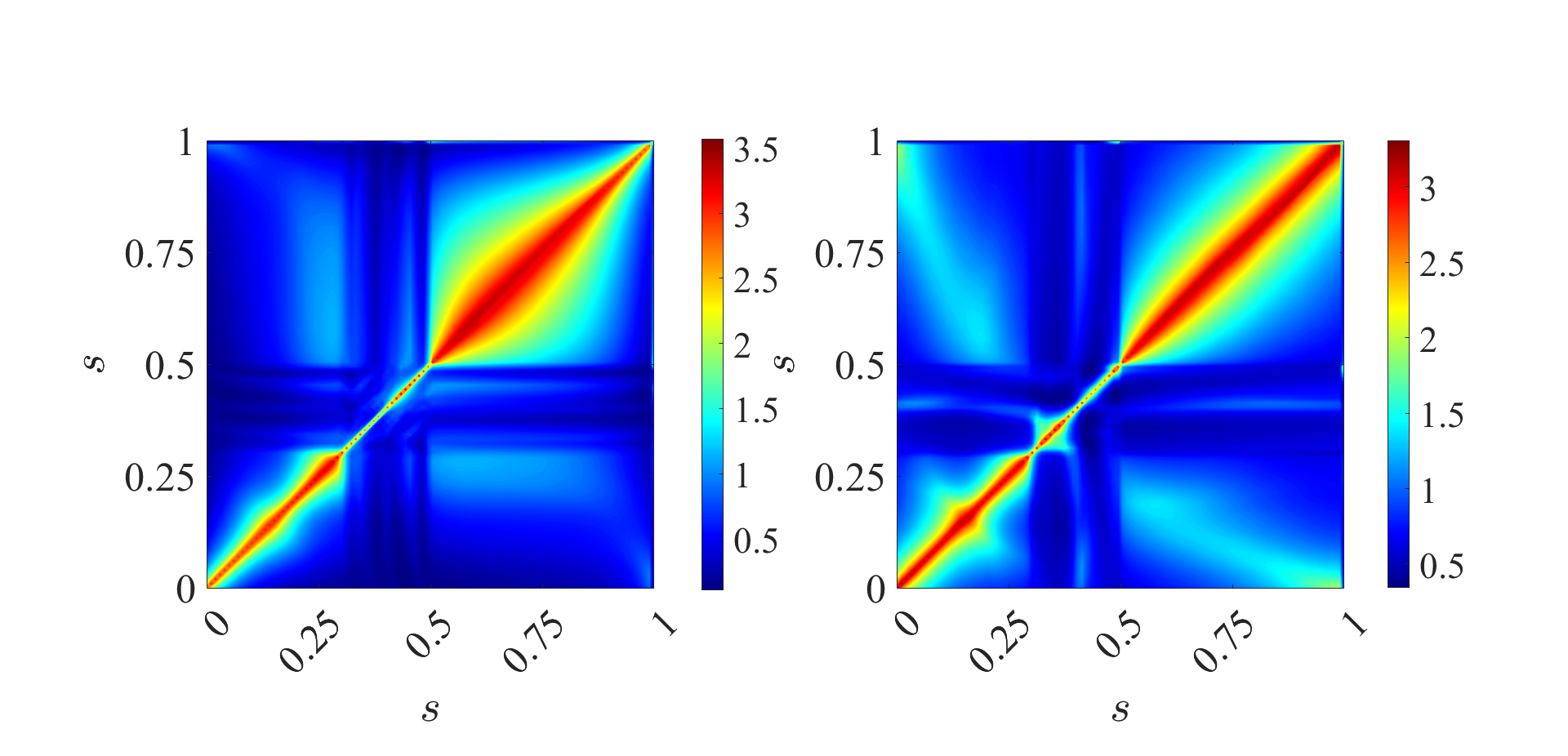}%
    \label{fig:P_MI_map}
}
\caption{Pressure covariance matrix cov$\left(P(s),P(s')\right)$ (a) and mutual information matrix $I\left(P(s),P(s')\right)$ (b) . Raw pressure, $P(s,t)$ (left), filtered pressure, $\tilde{P}(s,t)$ (right). Here $s$ is the generalized arc length around the airfoil measured clockwise from the leading edge (Fig \ref{fig:airfoil}).}
\label{fig:P_map}
\end{figure}

To investigate the spatial dynamics of the surface pressure we compute the covariance and mutual information matrices: $\operatorname{cov}\left(P(s,t),P(s',t)\right)$ and $I\left(P(s,t),P(s',t)\right)$ for both the raw pressure $P$ and the filtered pressure $\tilde{P}$. These quantify the information shared between different locations along the airfoil.  The covariance matrices and mutual information matrices are shown in figures \ref{fig:P_COV_map} and \ref{fig:P_MI_map} respectively. The left plot shows the raw pressure signal and the right shows the filtered pressure. We notice that the results for the raw and filtered pressure are qualitatively similar, and thus the following discussion applies to both.

These results reveal three distinct regions. First, the underside of the airfoil, $0.5<s<1.0$. This region displays a high degree of mutual information and strong correlation.
Second, the front section of the upper surface, $0<s<0.3$. This region exhibits similar features as the underside: strong mutual information and correlation, however, in this region the mutual information drops off much more quickly with separation between the sensor locations. These results imply that these regions are amenable to sparse sensor distribution, since any additional sensor is unlikely to contribute new information.  Note also that due to the airfoil having a non-zero angle of attack we see strong negative correlation between the upper and lower surface pressures. However, there is little mutual information between the upper and lower surfaces. This implies that measurements on one surface do not necessarily provide information about the other. The exception to this is the rear part of the upper surface, $0.3<s<0.5$. In this region there is little to no mutual information and significant variation in the correlation. As a result, this region likely requires relatively higher sensor density. We note that the transition point between the first two regions, $s=0.3$, coincides with the point of flow separation (see figure \ref{fig:airfoil}). Therefore the increased disorder observed in section 3 is likely due to to the complexity and increased unsteadiness of the flow in this region.

\begin{figure}
    \centering
    \includegraphics[trim = 110 0 0 0, scale = 0.30]{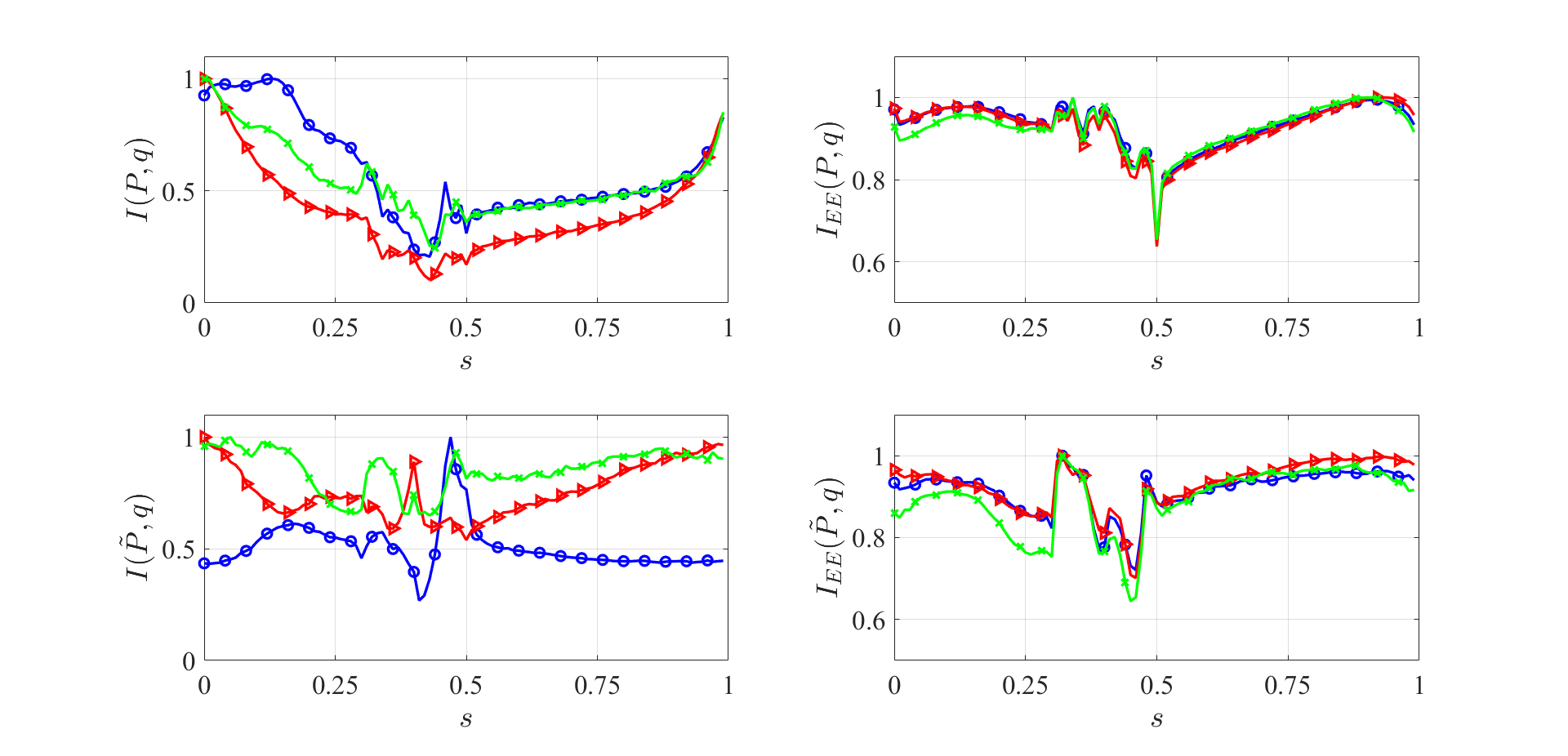}
    \caption{Mutual information (left) and extreme event mutual information (right) between raw (top row) and filtered (bottom row) pressure signal along the airfoil surface and the output drag coefficient for $\tau = 0,3,7$ (blue circles, red triangles, green crosses) as a function of arc length $s$. All curves are normalized by their maximum value.}
    \label{fig:Pq_MI_map}
\end{figure}

We also compute the standard and extreme event mutual information between the surface pressure and the drag coefficient: $I\left(P(s,t),q(t+\tau)\right)$ and $I_{EE}\left(P(s,t),q(t+\tau)\right)$. These are plotted in figure \ref{fig:Pq_MI_map} for the raw and filtered pressure signals for a range of $\tau$. As we are interested in the spatial variation of these quantities, to ease comparison we normalize each by its maximum value. In all cases we do not observe strong dependence on the lead time $\tau$.  As with the intra-pressure mutual information we see strong spatial dependence in the region $0.3<s<0.5$. For both the raw and filtered pressure signal, the extreme event mutual information is (locally) peaked in this region. On the other hand, for the standard mutual information this region lies in the trough of the spatial distribution. This suggests that the mechanisms driving the extreme events are strongest in the separation region. However, due to their rarity, this connection is not reflected in the standard mutual information profile. Next, we analyze this extreme event mechanism, and its connection to the extreme event mutual information in more detail.


\subsection{Extreme Event Mechanisms}\label{sec:mechanisms}
Extreme events occurring in dynamical systems are known to arise due to a variety of factors -- not all of which are fully understood. One class of dynamical system known to give rise to extreme events are slow-fast multi-scale systems \citep{Farazmand2019}. In such cases, the system evolves on two or more manifolds which have significant separation of characteristic time scales. At most times, the system evolves along the slower manifold. Occasionally, the trajectory may encounter an instability of this slow manifold, resulting in the trajectory rapidly approaching the fast manifold. Once the unstable region has passed the system relaxes back to the slow manifold. Such phenomena are often observed as sporadic high amplitude bursts \citep{Farazmand2019}. 

\begin{figure}
    \centering
    \includegraphics[trim = 30 0 0 0, scale = 0.3]{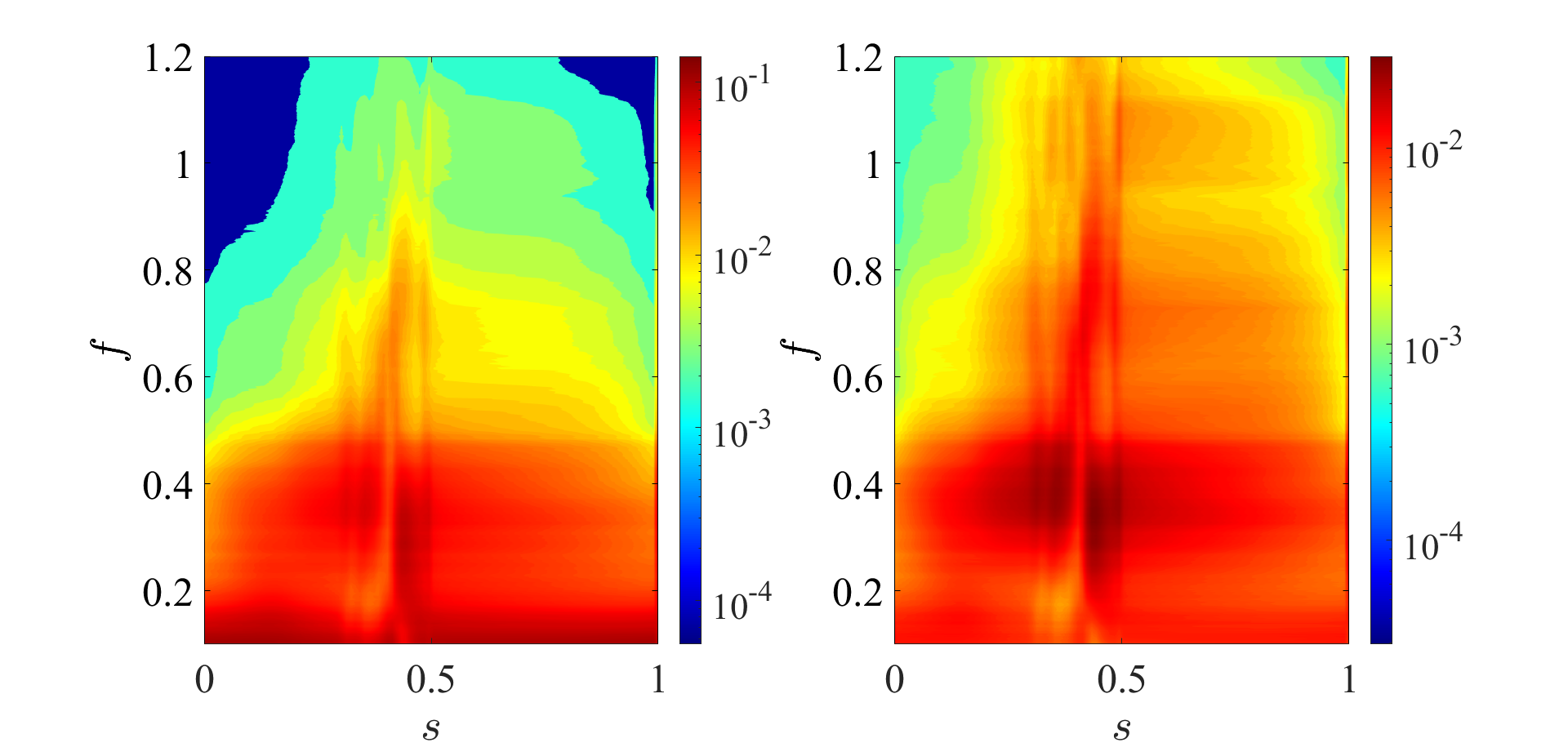}
    \caption{Standard (left) and pre-multiplied (right) power spectrum of filtered surface pressure. }
    \label{fig:fft_P}
\end{figure}

Airfoil flow is an example of such a multi-scale system. Such systems have multi-peaked spectral content -- or in other words they have multiple characteristic frequencies. At this Reynolds number there are two high frequencies (fast manifolds), the vortex shedding frequency, $f_{v} = 1.44$ corresponding to the energetically dominant oscillatory flow, and a second frequency corresponding to the extreme event manifold, $f_{e} = 0.4$. Figure \ref{fig:fft_P} shows the standard and pre-multiplied temporal Fourier power spectrum of the filtered surface pressure defined as
\begin{equation}
    \mathcal{P}(s,f) \equiv  \int P(s,t) e^{-ift}dt,
\end{equation}
and
\begin{equation}
    \mathcal{P}_{pm}(s,f) \equiv  f\mathcal{P}(s,f),
\end{equation}
respectively. 
The latter is useful for visualizing higher frequency content as it de-emphasizes the slow dynamics ($f\rightarrow0$).
In the standard power spectrum there is a clear maximum close to $f=0$, corresponding to the slow dynamics. The extreme event frequency (manifold) is also evident in the plain spectrum, but is best seen in pre-multiplied spectrum, which exhibits a clear peak around $f = 0.4$. We show the spectrum of the filtered pressure as the vortex shedding frequency at $f = 1.44$ is much stronger than either the slow or extreme event dynamics and obscures these when included in the spectrum. Note also the increased magnitude in the region $0.3<s<0.5$ consistent with the results of \S\ref{sec:MIstruct}.

\begin{figure}
    \centering
    \includegraphics[trim = 90 0 0 0, scale = 0.30]{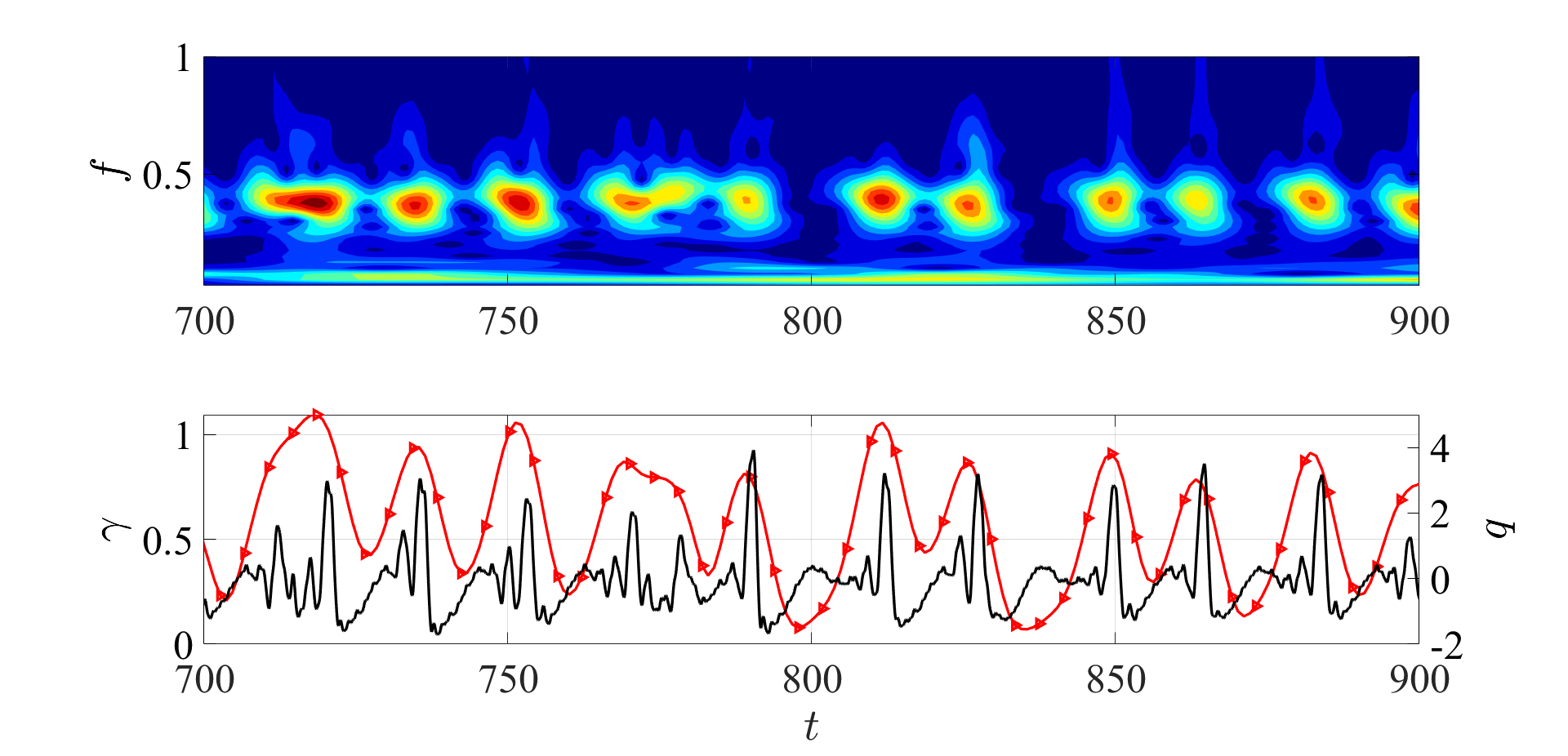}
    \caption{Absolute value of wavelet transformed pressure signal at $s = 0.34$ (upper panel). Extreme event indicator $\boldsymbol{\gamma} = |\hat{P}|_{f=f_{e}}$ at same location (red triangles) compared to drag coefficient (black line) (lower panel). }
    \label{fig:wavelet_P}
\end{figure}

The connection between this frequency and the extreme events is best interpreted through the wavelet transform. The wavelet transform allows for the visualization of the time varying strength of a signal's frequency content. The wavelet transform has been used in the past to identify extreme events by for example \citet{wavelet_sri} to detect bursts in pipe flow and \citet{cousins_sapsis, cousinsSapsis2015_JFM} for the early detection of rogue waves. The continuous wavelet transform (CWT) of a signal $x(t)$ is defined as
\begin{equation}
    \hat{X}(f,t) \equiv \mathcal{W}\left(x(t)\right)= \sqrt{\frac{f}{f_c}} \int^{\infty}_{-\infty} x(s) \psi\left(f \frac{s-t}{f_c}\right) ds.
\end{equation}
Here $\psi(t)$, is the wavelet function, and $f_c$ is the wavelet specific center frequency. The wavelet function is not unique, but must satisfy several conditions including finite energy and localized support \citep{addison_illustrated_2016}. Here we use the Morlet wavelet,
\begin{equation}
    \psi(t) = e^{-t^2/2}\cos(5t).
\end{equation}
Moving forward we refer to the wavelet transform of the pressure signal as $\hat{P}(s,f,t)$ where $f$ is the frequency. The wavelet transform of the pressure signal at $s = 0.34$ is shown in the upper panel of figure \ref{fig:wavelet_P}. This location corresponds to the peak in the spatial power spectrum in figure \ref{fig:fft_P}. Figure \ref{fig:wavelet_P} clearly shows the bursts of energy at $f = f_{e} = 0.4$.

\begin{figure}
    \centering
    \includegraphics[trim = 70 0 0 0, scale = 0.3]{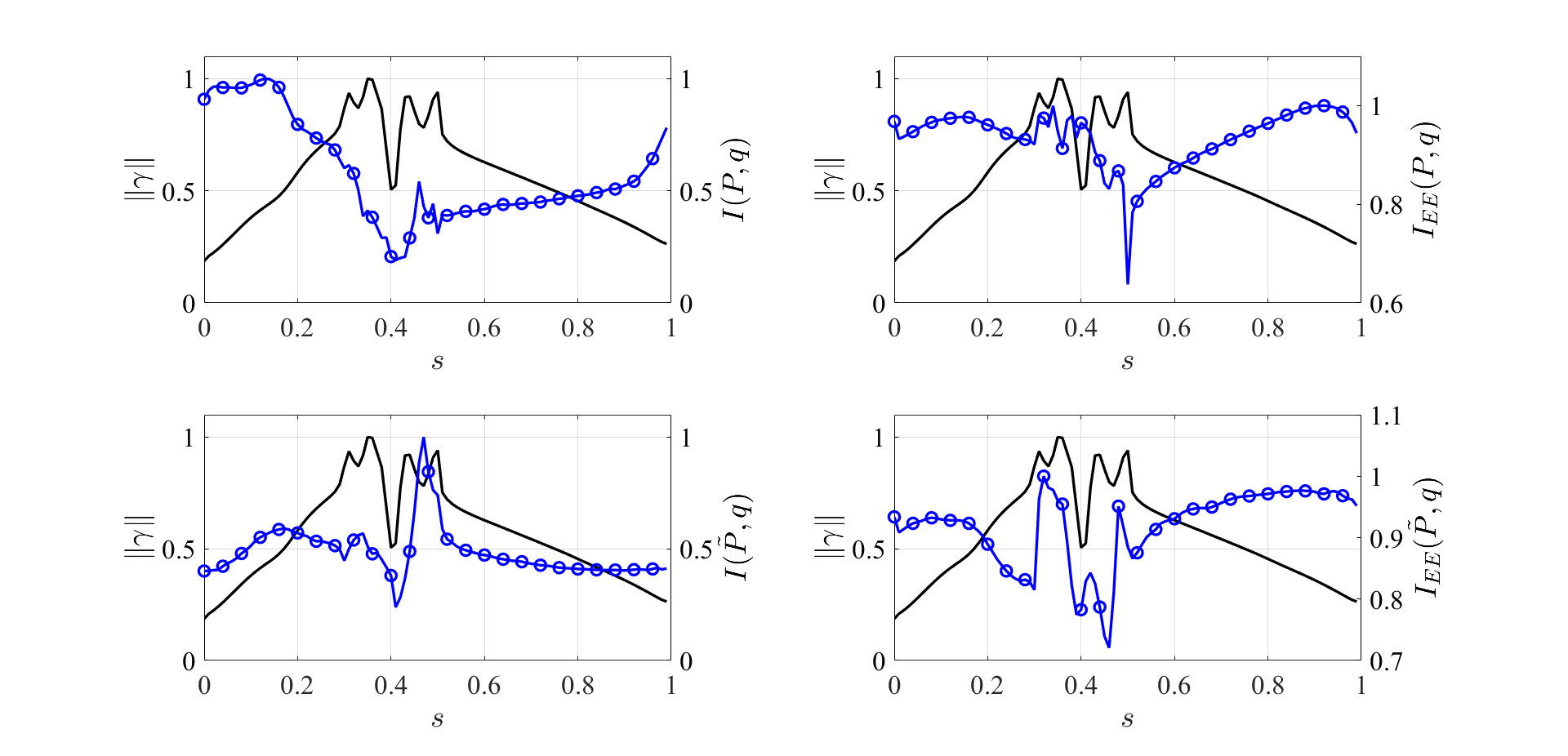}
    \caption{Spatial distribution of norm of wavelet coefficient evaluated at extreme event frequency (black line) compared to standard and extreme event mutual information between raw and filtered pressure signal and drag (blue circles). Raw pressure (top row), filtered pressure (bottom row), standard mutual information (left column), extreme event mutual information (right column). Blue lines are the same as in figure \ref{fig:Pq_MI_map}. }
    \label{fig:wavelet_MI}
\end{figure}

We define the extreme event indicator $\boldsymbol{\gamma}$ as the wavelet coefficient which maximizes the spectrogram of the filtered pressure signal, i.e. for $f=f_e$
\begin{equation}\label{gamma}
    \gamma(s,t) \equiv|\hat{P}(s,f_e,t)|.
\end{equation}
In the lower panel of figure \ref{fig:wavelet_P} we show the clear correlation between $\boldsymbol{\gamma}$ and the extreme drag events. This connection is even further highlighted in figure \ref{fig:wavelet_MI} where we compare the spatial dependence of the norm 
\begin{equation}
    \|\gamma(s)\| =\sqrt{ \int |\gamma(s,t)|^2 dt},
\end{equation}
to the standard and extreme mutual information profiles (also shown in figure \ref{fig:Pq_MI_map}). As we previously found that these mutual information profiles do not depend significantly on $\tau$, we show only the distribution for $\tau = 0$. Notice that the norm of the wavelet coefficient peaks in the same region of the airfoil, $0.3<s<0.5$, as the extreme event mutual information between the pressure signal and the drag. This suggests that the extreme events observed in this flow are indeed of the multi-scale system type, and that the connection (as quantified by the extreme event mutual information) is a reflection of the strength of the extreme event frequency. In \S\ref{sec:wavelet} we show that preprocessing the pressure signal to extract this extreme event manifold allows for a drastic reduction in the model complexity and data required for accurate forecasting.




\subsection{Flowfield Analysis}\label{sec:flowfield}
In \S \ref{sec:MIstruct} and \S\ref{sec:mechanisms} we identified the unstable extreme event manifold and its connection to the separation region of the airfoil. Here we zoom out and analyze the full boundary layer to better understand the mechanisms at the heart of the instability and the subsequent extreme drag events. To this end we follow the time evolution of the vorticity field, 
\begin{equation}
    \omega(x,y,t) \equiv  \frac{\partial v}{\partial x} -\frac{\partial u}{\partial y} 
\end{equation}
over the course of a single extreme event from $t=911-931$, with a peak at $t = 921$. Figure \ref{fig:vort_snaps} shows 8 snapshots of the vorticity field over this time interval -- for clarity we focus on the region near the boundary layer. The corresponding values of the instantaneous drag coefficient are shown in the top left panel of the same figure. The red markers in the latter represent the time instances of the 8 vorticity snapshots.  We see clear evidence of boundary layer separation/disorder during the time instance corresponding to the peak in drag coefficient. Although we show only a single extreme event here, this disordered behaviour of the boundary layer was observed during the peak of all extreme events. 

The transient dynamics observed in the full vorticity field are subtle, and thus to better understand these transient dynamics we compute the wavelet transform of the entire vorticity field,
\begin{equation}
    \hat{\Omega}(x,y,t,f) = \mathcal{W}\left( \omega(x,y,t) \right).
\end{equation}
This allows us to investigate the component of the vorticity evolving with the extreme event frequency identified in \S\ref{sec:mechanisms} -- which is not the dominant energetic contributor to the full field, and is thus liable to be obscured in the snapshots in figure \ref{fig:vort_snaps}. 

\subsubsection{Global Dynamics}
\begin{figure}
    \centering
    \includegraphics[trim = 190 0 0 0, scale = 0.19]{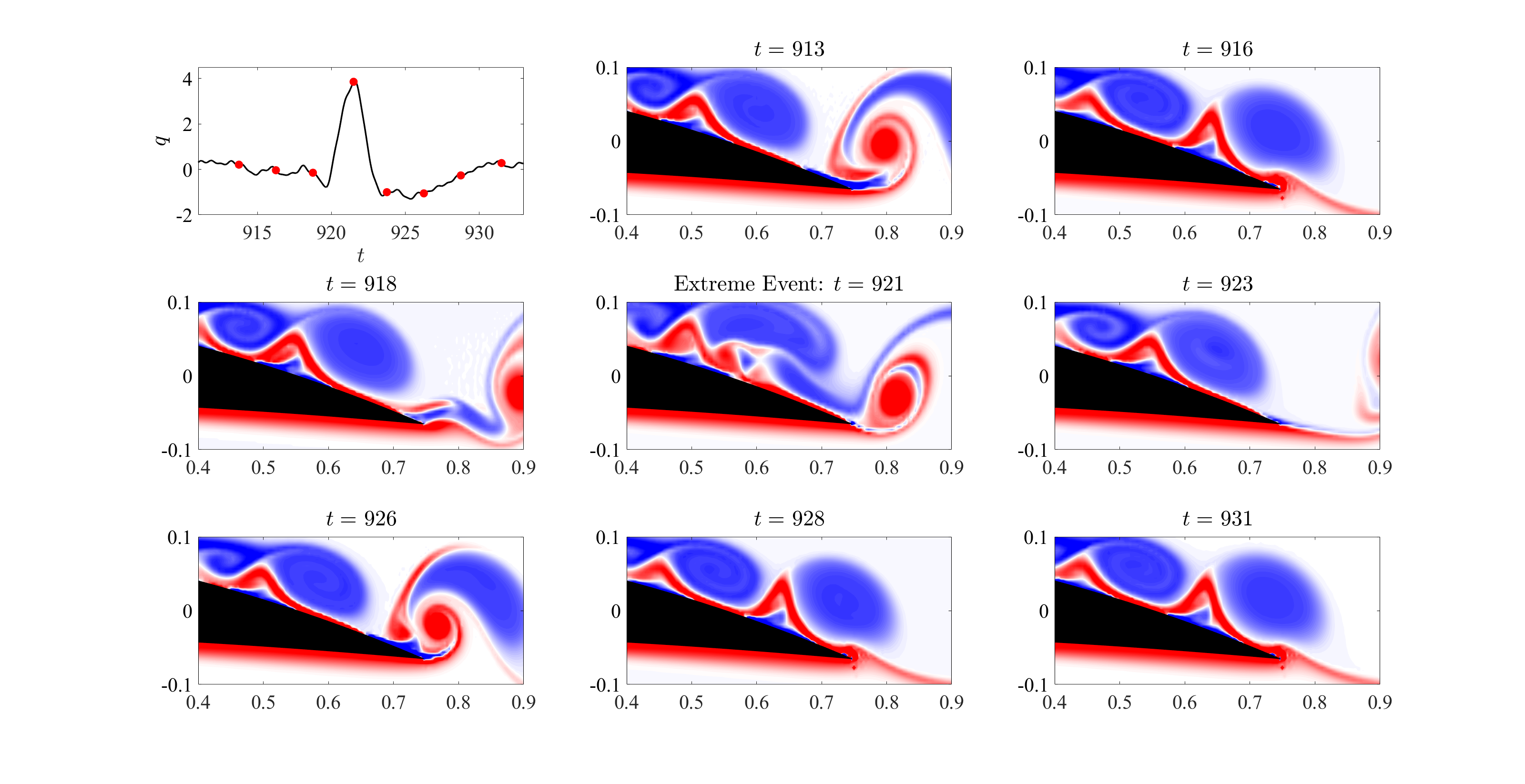}
    \caption{Drag coefficient (top left) and temporal snapshots of vorticity field traversing a single extreme event. Red markers in top left plot indicate snapshot time instances. }
    \label{fig:vort_snaps}
\end{figure}

To analyze the dynamics of the boundary layer we first consider the wavelet transform of the full flowfield -- focusing on the the extreme event frequency, 
\begin{equation}\label{Omega_e}
   \hat{\Omega}_{e}(x,y,t) \equiv \hat{\Omega}(x,y,t,f)|_{f=f_e=0.4},
\end{equation}
and the vortex shedding frequency,
\begin{equation}\label{Omega_v}
   \hat{\Omega}_{v}(x,y,t) \equiv \hat{\Omega}(x,y,t,f)|_{f=f_v=1.44}.
\end{equation}
The wavelet component associated with the extreme event frequency (\ref{Omega_e}) and the vortex shedding frequency (\ref{Omega_e}) are plotted in figures \ref{fig:wavelet_snaps} and \ref{fig:wavelet_snaps_vs} respectively for the same time instances as in figure \ref{fig:vort_snaps}. In the former, we see clear evidence of a coherent structure with relatively small characteristic spatial length scale which undergoes a transient instability resulting in a temporary loss of coherence during the extreme event before recovering as the drag coefficient returns to its nominal state. The vortex shedding mode has a much larger characteristic length scale -- on the order of the vortical structures seen in figure \ref{fig:vort_snaps} -- and does not appear to undergo any significant changes during the extreme event. 



To further illustrate the dynamics of these two frequency components we compute the temporal correlation function
\begin{equation}\label{vort_proj}
    r_{\alpha}(t_0,t) \equiv \left|\int\int \hat{\Omega}^*_{\alpha}(x,y,t_0) \hat{\Omega}_{\alpha}(x,y,t) dx dy\right|, 
\end{equation}
where $^*$ denotes the complex conjugate and $\alpha = e,v$ and the integration is performed over the entire domain. For the special case where $t_0=t$ this is equivalent to the $L_2$ norm of the wavelet mode
\begin{equation}\label{vort_norm}
    \|\hat{\Omega}_{\alpha}\|^2(t) \equiv \int\int |\hat{\Omega}_{\alpha}(x,y,t)|^2 dx dy.
\end{equation}
These metrics respectively quantify the temporal evolution of the shape (length scale) and magnitude of the vorticity at a specific temporal frequency.

The correlation (\ref{vort_proj}) is plotted in the upper panel of figure \ref{fig:wavelet_norms} for both $\hat{\Omega}_{e}$ (blue circles) and $\hat{\Omega}_{v}$ (red triangles). We fix $t_0 = 911$ as a representative snapshot corresponding to the vorticity structure during the quiescent periods -- however, any quiescent time instance could be used. We clearly observe a systematic and drastic loss of coherence in the extreme event mode during the spikes in the drag coefficient. Inspection of figure \ref{fig:wavelet_snaps} suggests that this is at least in part due to an increase in the dominant spatial length scale. The coherence of the vortex shedding mode actually fluctuates -- with what further analysis reveals to be at close to the extreme event frequency -- with an amplitude that slightly increases during the extreme events, however no drastic loss of coherence is observed. The significance of this fluctuation is not immediately clear, but it suggests some interaction between the two frequency components -- the investigation of which is the focus of ongoing research.

The central and lower panel of figure \ref{fig:wavelet_norms} show the time evolution of the norm (\ref{vort_norm}) of the extreme event mode (blue circles) and vortex shedding mode (red triangles). The central panel shows the large difference in magnitude between these two frequency components -- the vortex shedding mode generally contains an order of magnitude more energy than the extreme event mode. The lower panel compares these norms (normalized by their value at $t_0$) to the drag coefficient (black). For reference, in the lower panel we also plot the normalized surface pressure extreme event wavelet coefficient, $\gamma(s,t)_{s=0.34}$ (\ref{gamma}) (green squares). As in the previous sections, the location $s = 0.34$ is chosen as it is located within the separation region. While as previously noted, the magnitude of $\gamma$ peaks in sync with the extreme drag events, the global norm of the extreme event component of the vorticity, $\hat{\Omega}_e$, drops in magnitude during the same time intervals. In contrast to both of these, the dominant vortex shedding mode, (\ref{Omega_v}), is significantly more stable and exhibits a much smaller relative drop in magnitude during the extreme events.

\subsubsection{Dynamics on the Airfoil Surface}
\begin{figure}
    \centering
    \includegraphics[trim = 190 0 0 0, scale = 0.19]{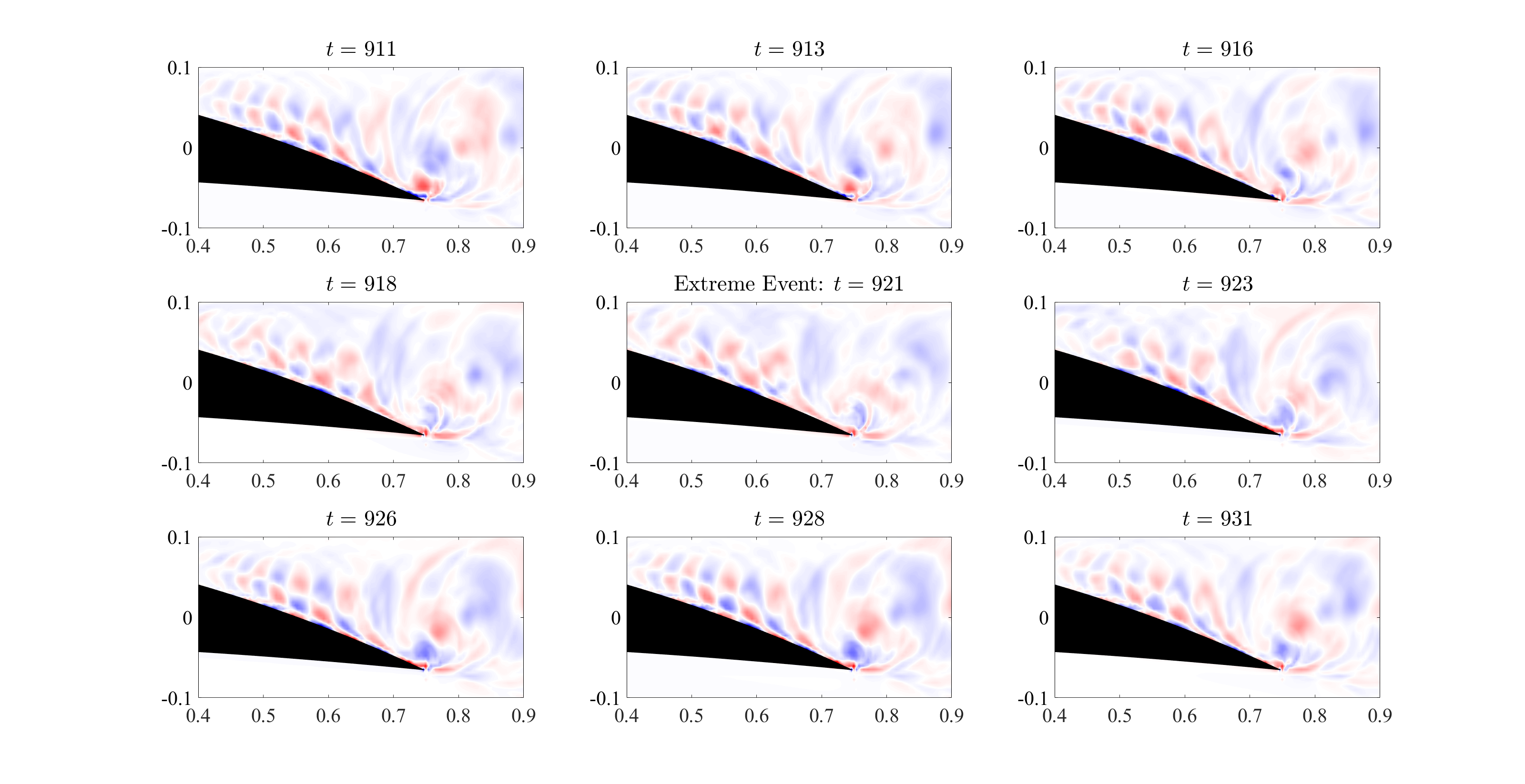}
    \caption{Temporal snapshots of the vorticity wavelet coefficient evaluated at the extreme event frequency $\hat{\Omega}_{e}(x,y,t)$. Same time snapshots as figure \ref{fig:vort_snaps}. Note the loss of coherence and increase in spatial scales during the extreme event.}
    \label{fig:wavelet_snaps}
\end{figure}

\begin{figure}
    \centering
    \includegraphics[trim = 190 0 0 0, scale = 0.19]{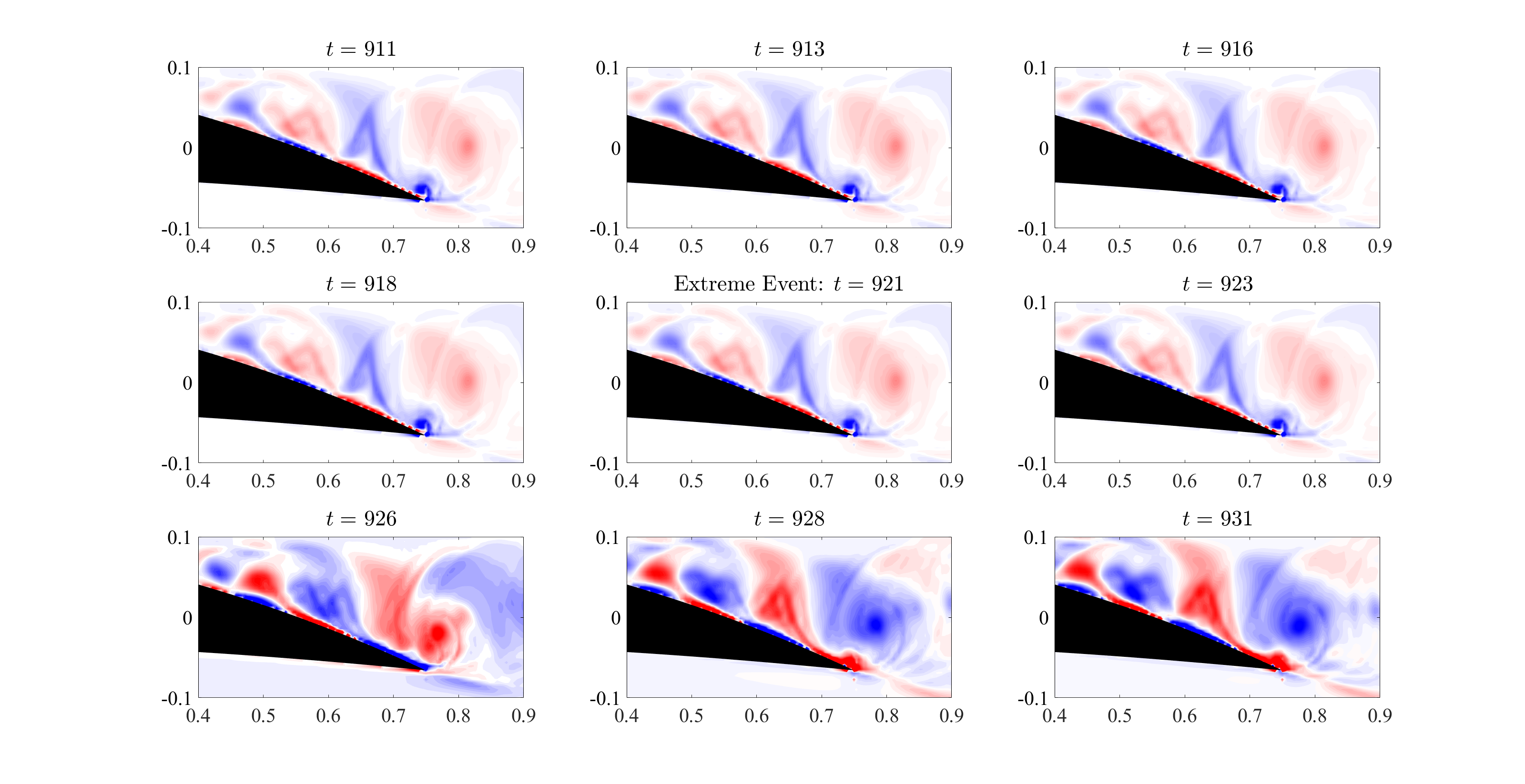}
    \caption{Temporal snapshots of the vorticity wavelet coefficient evaluated at the vortex shedding frequency $\hat{\Omega}_{v}(x,y,t)$. Same time snapshots as figures \ref{fig:vort_snaps} and \ref{fig:wavelet_snaps}. No significany spatial changes are observed during the extreme event, indicating that vortex shedding mode acts as energy reservoir.}
    \label{fig:wavelet_snaps_vs}
\end{figure}

\begin{figure}
    \centering
    \includegraphics[trim = 230 0 0 0, scale = 0.20]{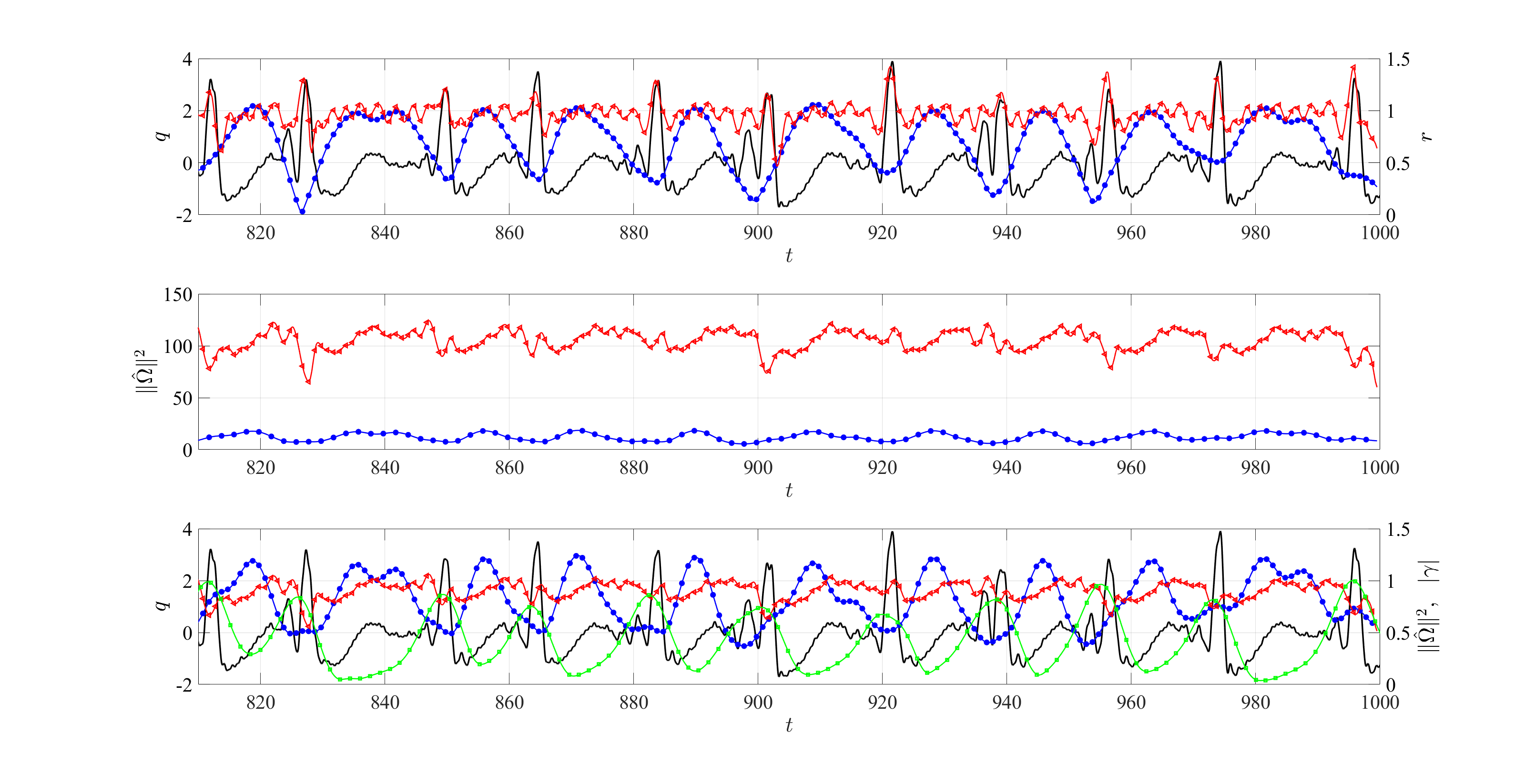}
    \caption{Top panel: normalized correlation function (\ref{vort_proj}) for $t_0 = 911$ for $\hat{\Omega}_{e}$ (blue circles) and $\hat{\Omega}_{v}$(red triangles). Center panel: vorticity wavelet coefficient norms $\|\hat{\Omega}_{e}\|^2$ (blue circles), $\|\hat{\Omega}_{v}\|^2$ (red triangles). Lower panel: normalized vorticity wavelet coefficient norms (same markers as central panel) and normalized surface pressure wavelet coefficient corresponding to extreme event frequency, $\gamma(s,t)_{s=0.34}$ (\ref{gamma}) (green squares). In the upper and lower panel the drag coefficient $q(t)$ is shown in black.   }
    \label{fig:wavelet_norms}
\end{figure}

\begin{figure}
    \centering
    \includegraphics[trim = 50 0 0 0, scale = 0.26]{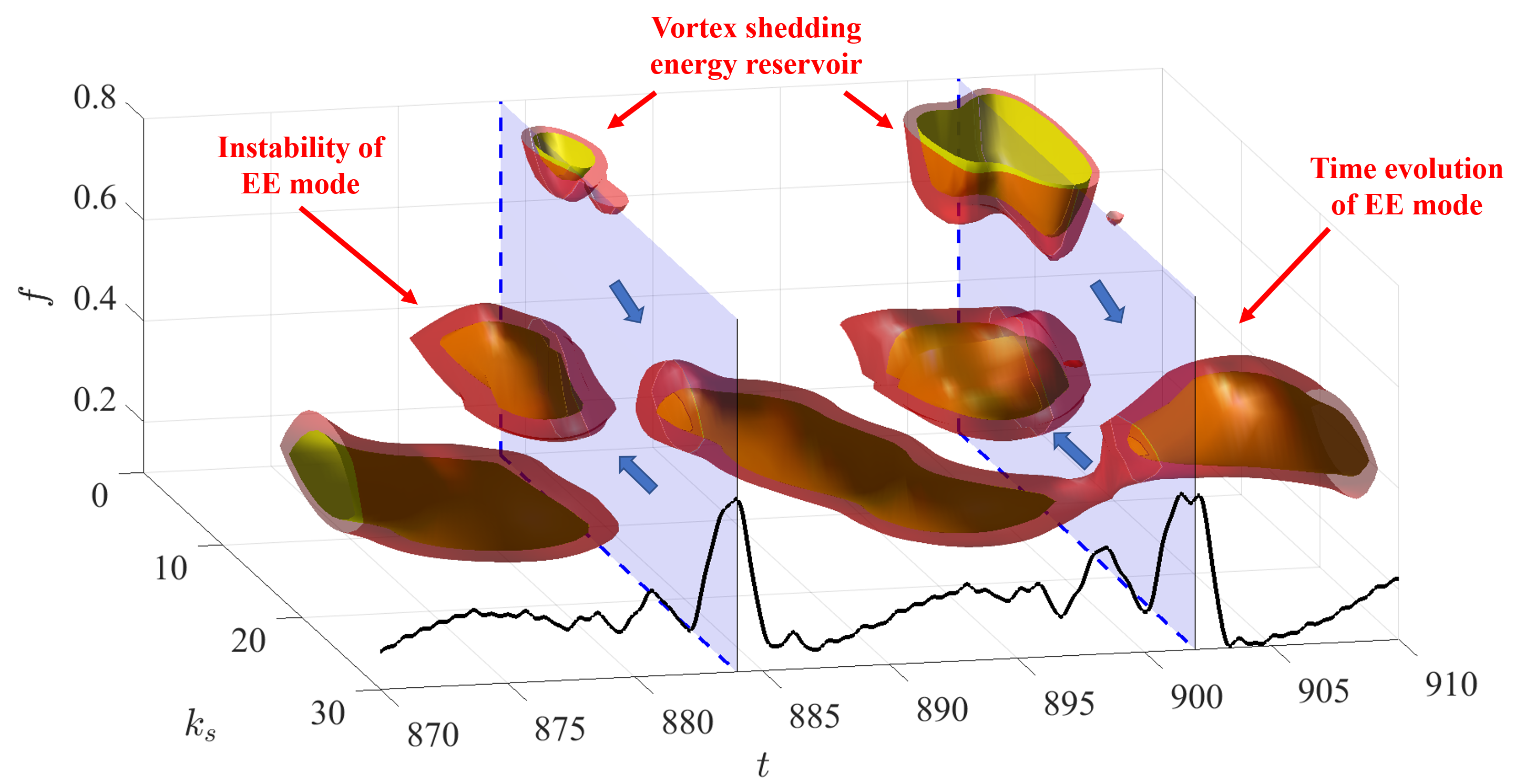}
    \caption{Isocontours of vorticity amplitude on the airfoil surface as a function of temporal frequency, $f$, spatial wavenumber, $k_s$, and time, $t$. Blue arrows represent direction of energy transfer.}
    \label{fig:spatial_fft}
\end{figure}

To better understand the instability of the extreme event mode and the associated transfer of energy we also compute the spatial Fourier transform of the temporally wavelet transformed vorticity evaluated at the airfoil surface. In other words we compute 
\begin{equation}
    \tilde{\Omega}_s(t,f,k_s) \equiv \left| \mathcal{F}\left( \hat{\Omega}(x(s),y(s),t,f)  \right)\right| = \left|\int \hat{\Omega}(x(s),y(s),t,f) e^{-ik_s s}ds\right|
\end{equation}
where $k_s$ is the spatial wavenumber with respect to the arclength $s$ along the airfoil surface defined in \S\ref{sec:problem}  and $x(s)$ and $y(s)$ are the coordinates of that surface. This reduces the spatial dimensions from two to one, and thus allows us to visualize the transfer of energy between various spatial and temporal scales as a function of time. The isocontours of this quantity are plotted in figure \ref{fig:spatial_fft} over a time horizon covering two extreme events -- the drag coefficient is also plotted for reference. This plot succinctly summarizes the observations discussed above. First we see that during the quiescent periods, the energy of the extreme event mode, which actually seems to meander slightly about $f=0.4$, is concentrated at a wavenumber $k_s \approx 20$. Then during the extreme events, energy is drawn from the higher frequency vortex shedding mode, which serves as an energy reservoir, leading to the instability of the extreme event mode which abruptly transfer its energy to a lower wave number $k_s \approx 10$. This appears in figure \ref{fig:spatial_fft} as the \textit{``pinching off''} of the isocontours during the spikes in the drag. This is the manifestation of loss of coherence and increase in spatial scale of the extreme event mode observed in figure \ref{fig:wavelet_snaps} and quantified in the upper panel of \ref{fig:wavelet_norms}. We note in closing that this phenomenon of the extreme events evolving on a manifold distinct from the energetically dominant one has been observed in other systems such as for example Kolmogorov flow \citep{farazmand_variational_2017,wan_data-assisted_2018}. In that case the energy of the flow is dominated by a specific triad of spatial wavenumbers, however projecting the flow onto this triad fails to predict the extreme energy dissipation events observed in that flow. Similarly, in the case of the airfoil flow considered here, extracting only the dominant vortex shedding dynamics would miss the extreme event dynamics entirely. However, our findings are contrary to the far more common phenomenon of instabilities transferring energy from large to small scales.




\FloatBarrier
\section{Offline Sparse Sensor Placement}\label{sec:offline}
The analysis of mutual information structure described in \S\ref{sec:stats} indicates that certain sections of the airfoil are statistically more informative of the drag coefficient. To test the practical implications of this discovery we first propose an offline strategy to optimally select sensor locations. Such an algorithm does not require actually training the neural network. Therefore, it can be thought of as a prepossessing step which allows us to optimally design the network architecture prior to training. At each iteration, the sampling algorithm, which is outlined in algorithm \ref{alg:a1}, selects the next best sensor location by maximizing a cost function referred to as an acquisition function. Throughout this work we use the term ``acquisition function'' strictly in connection with such a sampling strategy, and the term ``cost function'' to refer to the cost function used to train a given model.

For a given application, the choice of acquisition function is not obvious, see for example \citet{Chaloner95,sapsis20, Yang2021}. In this framework sensor locations are selected sequentially, and therefore we seek locations which are maximally informative of the drag coefficient and minimally redundant with respect to the previously placed sensors. Thus, we propose the following two acquisition functions based on the previously defined standard and extreme event mutual information
\begin{equation}\label{a1_MI}
   a^1_{j+1}(s,\textbf{s}^*_{j}) = \frac{I\left(P(s),q\right)}{\frac{1}{j}\sum_{k=1}^{j}I\left(P(s),P(s_k^*)\right)},
\end{equation}
\begin{equation}\label{a2_MI}
   a^2_{j+1}(s,\textbf{s}^*_{j}) = \frac{I_{EE}\left(P(s),q\right)}{\frac{1}{j}\sum_{k=1}^{j}I\left(P(s),P(s_k^*)\right)}.
\end{equation}
The numerator -- the mutual information between the pressure signal with the drag coefficient -- rewards predictive capability. The denominator -- the average of the intra-pressure sensor mutual information -- penalizes redundancy. This second condition ensures that sensors are not placed in locations which do not contribute information not already encoded in previously placed sensors. 

There is no unique way to quantify the connection between a prospective sensor location and the previously placed sensors, and the average used here is only one option. Therefore, we also considered a second acquisition function where the arithmetic mean in the denominator of (\ref{a1_MI}) and (\ref{a2_MI}) is replaced with a geometric mean, but we did not observe significantly different results. A more exhaustive study of candidate functions is beyond the scope of this work, and so for the sake of brevity we restrict ourselves to (\ref{a1_MI}) and (\ref{a2_MI}). Going forward we refer to any results obtained using this algorithm as offline-mutual-information ($OMI_N$) where $N$ is the number of sensors.

\begin{algorithm}
\caption{Optimal Placement of $N$ sensors}\label{alg:a1}
\begin{algorithmic}
\While{$j < N$}

$s^*_{j+1} = \operatorname{argmax} a_{j+1}(s,\textbf{s}^*_{j})$

update $\mathbf{s}_{j+1}^* = [\mathbf{s}^*_{j},s^*_{j+1} ]$

\EndWhile
\end{algorithmic}
\end{algorithm}

\subsection{Results: Sensor Placement}\label{sec:offline_locs}
\begin{figure}
    \centering
    \includegraphics[trim = 60 0 0 0,scale = 0.33]{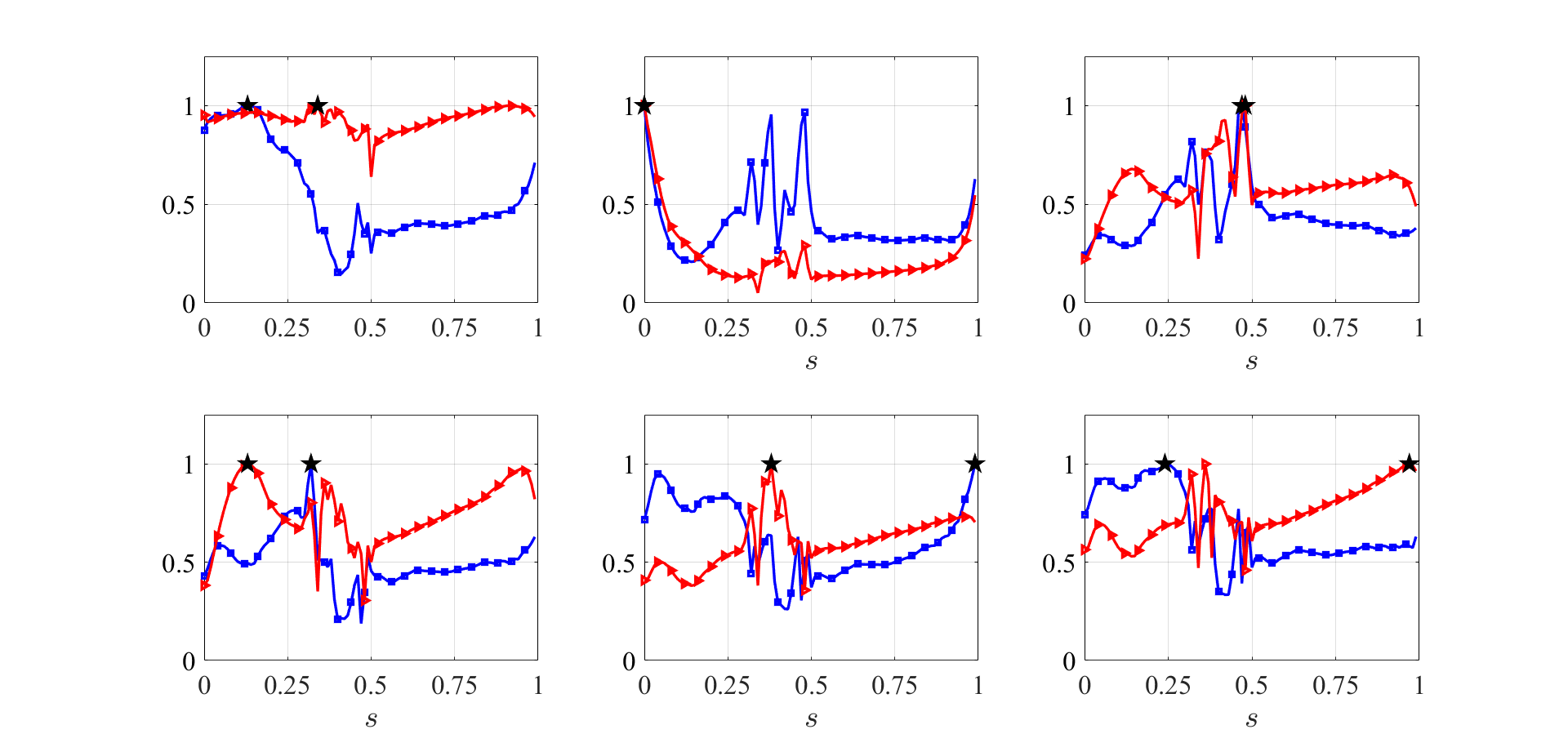}
    \caption{OMI acquisition function landscape for sensors 1-6 from left to right, top to bottom. Black stars indicates maximum point. Standard mutual information (blue squares), extreme event mutual information (red triangles).}
    \label{fig:omi_aq_fun}
\end{figure}

\begin{figure}
    \centering
    \subfloat[]{%
   \includegraphics[trim = 60 0 0 0,scale = 0.30]{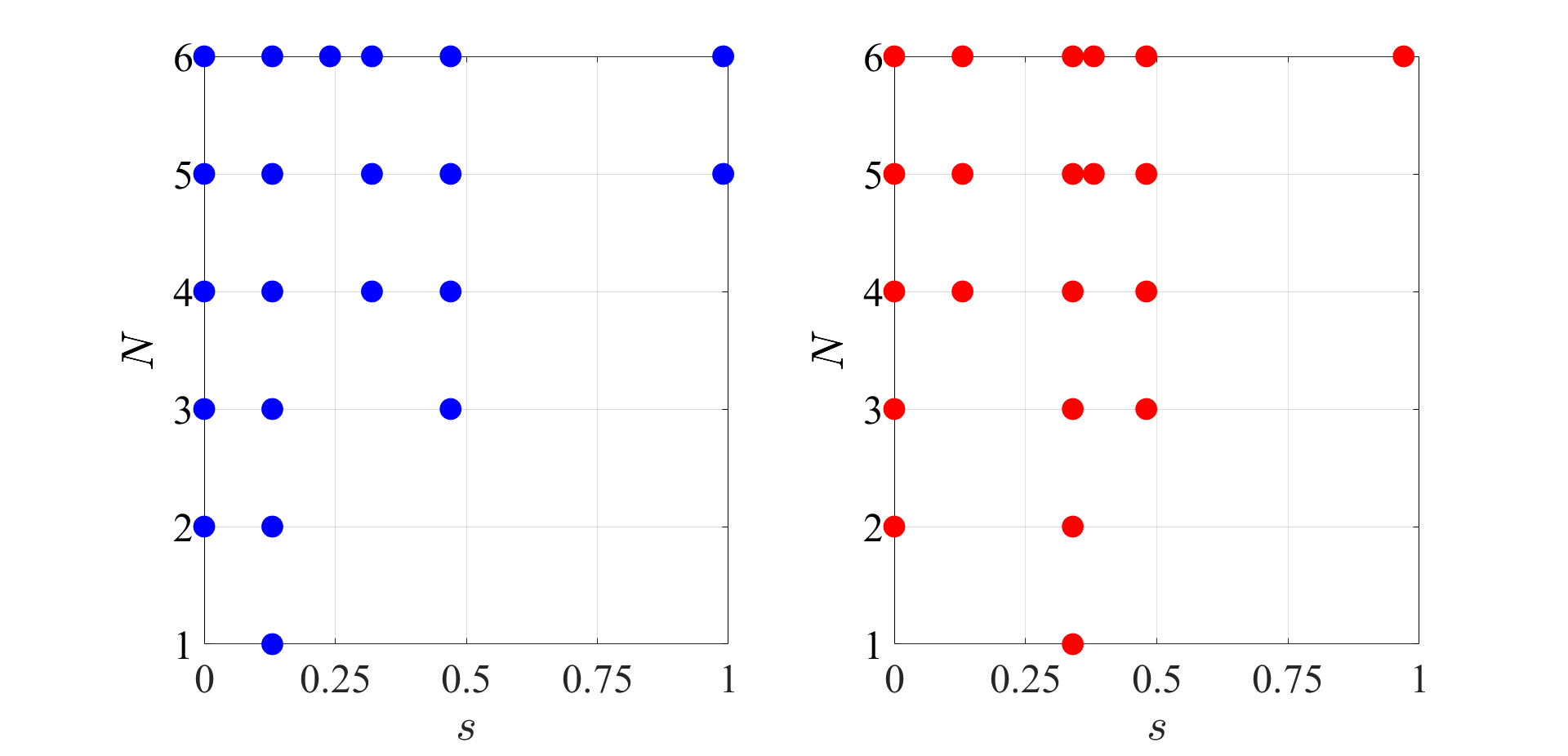}
    \label{fig:omi_sensors1}
}

\subfloat[]{%
   \includegraphics[trim = 30 0 0 0,scale = 0.30]{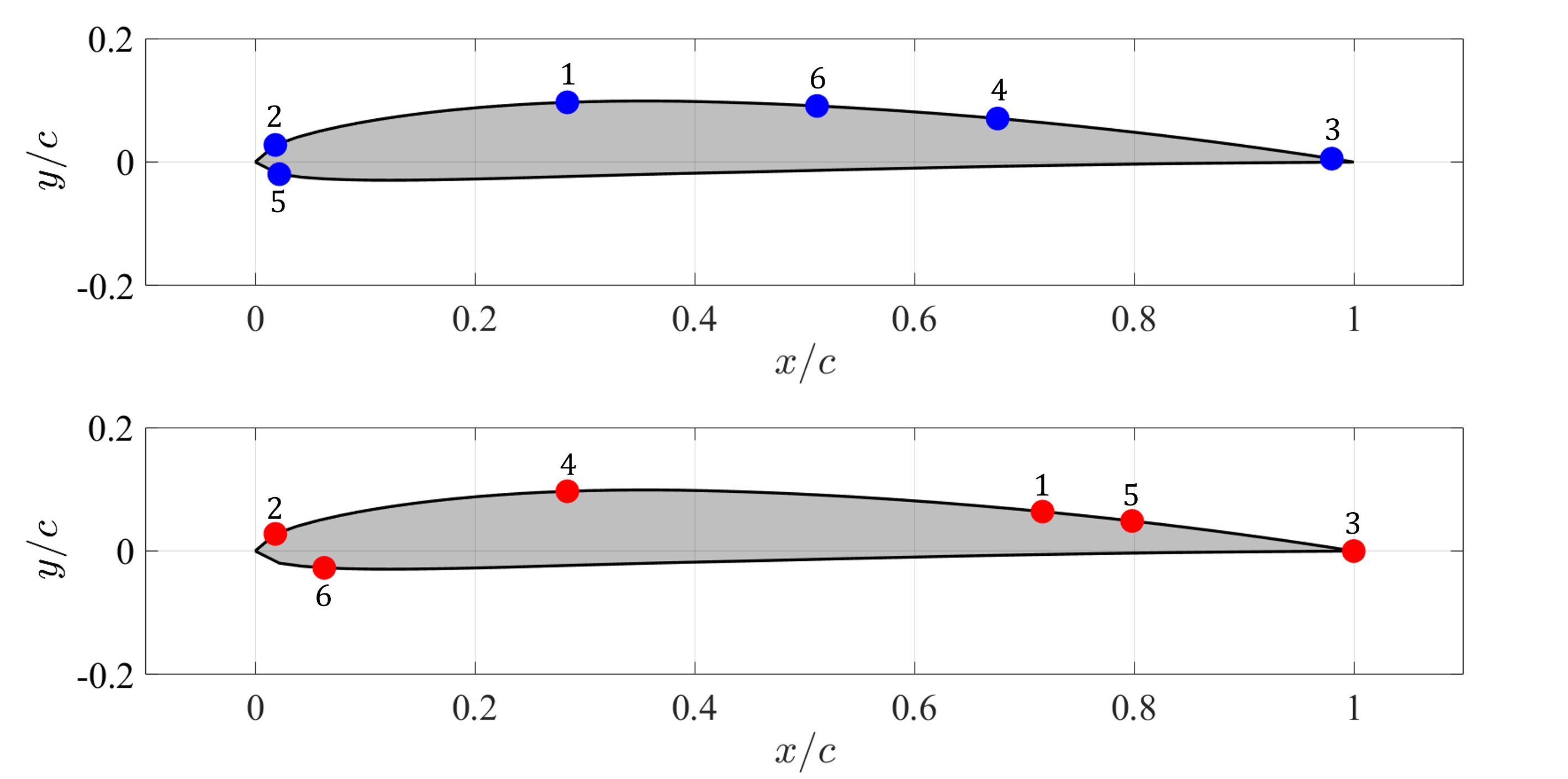}
   \label{fig:omi_sensors2}
}
 \caption{Optimal sensor locations after each iteration $N$.  Standard mutual information (left), extreme event mutual information (right) (a). Visualization of optimal sensor locations on airfoil, standard mutual information (blue), extreme event mutual information (red) (b). Sensor locations are labeled by $N$. }
    \label{fig:omi_sensors}
\end{figure}

We apply algorithm \ref{alg:a1} with acquisition functions (\ref{a1_MI}) and (\ref{a2_MI}) to our data set to compute the first six optimal sensor locations. Because the results of figure \ref{fig:P_MI_map} indicate that the general behaviour of the mutual information is not dependent on the lead time $\tau$, we fix $\tau = 0$. Additionally, in order to facilitate comparison with \citet{rudy_prediction_2022}, we consider only the raw pressure signal. The acquisition function landscape for at each iteration is plotted in figure \ref{fig:omi_aq_fun}.  The optimal senor locations after each iteration are then summarized in figure \ref{fig:omi_sensors}.

The globally optimal sensor locations are simply the points of maximum standard and extreme event mutual information -- these are located at approximately $s = 0.15$ for (\ref{a1_MI}) and $s= 0.3$ for (\ref{a2_MI}) respectively. However, inspection of figure \ref{fig:omi_aq_fun} indicates that in the latter case the acquisition function landscape does not display any significant variation along the airfoil, calling into question the viability of (\ref{a2_MI}) as a practical metric for optimal sensor placement. For the standard mutual information case, (\ref{a1_MI}), at iteration 2-4 the acquisition function exhibits multiple local maxima of roughly equal value in the region $0.3<s<0.5$. These multiple peaks are sequentially ``picked off'' throughout the iterations 2-4. This phenomenon is also observed, but to a slightly lesser extent, in the extreme event mutual information case (\ref{a2_MI}). The similarity of these local maxima mean that these 4 sensor locations should be thought of as an ``optimal grouping'' rather than a strict ranking, since measurement noise or numerical errors could affect their ordering. However, we note that adding small amounts of noise did not significantly impact the qualitative features of the results. Figure \ref{fig:omi_sensors} highlights that while the ordering of the sensors varies between the standard and extreme event versions of the model, the final distribution of the optimal sensors is qualitatively very similar.

The under side of the airfoil $(0.5<s<1)$ is completely ignored by the algorithm until iteration 5 for the standard case and iteration 6 for the extreme event case, where a strong maximum is observed just downstream of the leading edge. This solitary underside sensor near the leading edge is consistent with the mean pressure profile observed in the flow over an inclined airfoil. The mean pressure gradient (w.r.t. arc length) is generally significant along the upper surface but relatively weak along the lower surface. Therefore, a single sensor can capture a significant amount of the information of the pressure field along the lower surface, since once the jump in pressure across the leading edge is established there is not much more to be gained from further probing the pressure along the underside of the airfoil. 

\subsection{Results: Evaluation}\label{sec:offline_NN}
To test the efficacy of the proposed offline sensor placement algorithm we train the same LSTM network described in \citet{rudy_prediction_2022} using the first 5 optimal sensor locations predicted using algorithm \ref{alg:a1} with acquisition function (\ref{a1_MI}). Due to the similarities of the sensor locations predicted by (\ref{a1_MI}) and (\ref{a2_MI}) and the high computational cost of training the network we omit the predictions of (\ref{a2_MI}) from this analysis. We compare those results to those in \citet{rudy_prediction_2022} using 50 sensors spaced equally around the airfoil. The network architecture is 
\begin{equation}\label{rudy_net}
    \rightarrow FC32 \rightarrow LSTM32 \rightarrow LSTM32 \rightarrow FC32 \rightarrow FC16 \rightarrow FC8 \rightarrow FC4 \rightarrow FC1 \rightarrow
\end{equation}
where FC and LSTM stand for \textit{`fully-connected'} and \textit{`long-short term memory'} respectively, and the swish activation function is applied between each layer \citep{ramachandran_searching_2017}. In order to isolate the effects of the sensor placement, we make no changes to the architecture or other than the input dimension and utilize the same training strategies as \citet{rudy_prediction_2022}. Training is conducted using $70\%$ of the data, with the remaining 30 $\%$ split evenly between validation and testing. The model was trained over 3 restarts using 140 history points and a mean square error loss function until the validation error failed to decrease for 10 epochs -- no regularization was used. The interested reader is referred to \citet{rudy_prediction_2022} for a more detailed description of the network architecture and training strategy. 

We compare three different models: the reference case from \cite{rudy_prediction_2022} using 50 sensor locations, the $OMI_5$ model, as well as a second reference case using 5 uniformly spaced sensors -- all three use the raw pressure data as an input. The last case is included to verify that any potential benefit of our algorithm is actually due to the algorithmic placement of the sensors and not simply a reflection of oversampling by \cite{rudy_prediction_2022}. The three models are summarized in table \ref{tab:NN_models}.

To compare the predictive capabilities of the models we compute both the mean square error (MSE) of the model prediction as well as the \textit{maximum adjusted area under the precision-recall curve} -- a metric introduced  by \citet{guth_machine_2019} which quantifies the accuracy of extreme event prediction. The area under the precision-recall curve is then defined as
\begin{equation}
    \alpha(\chi)=\int_0^1 S(R,\chi) dR,
\end{equation}
where the event rate $\chi$ is defined as the probability that the output exceeds some threshold, the precision, $S$, the ratio of correct event predictions to total event predictions, and the recall, $R$, is the ratio of correct event predictions to the actual number of events. The maximum adjusted value is then defined as
\begin{equation}\label{guth_crit}
    \alpha^* = \operatorname{max}_{\chi} \left(\alpha(\chi)-\chi\right).
\end{equation}
When the value of $\alpha^*$ is large (approaches unity) the model is very good at predicting rare events, alternatively, when the value approaches zero a model does no better than a guess based on the aggregate frequency of extreme events. 

\begin{figure}
    \centering
    \includegraphics[trim = 80 0 0 0, scale = 0.33]{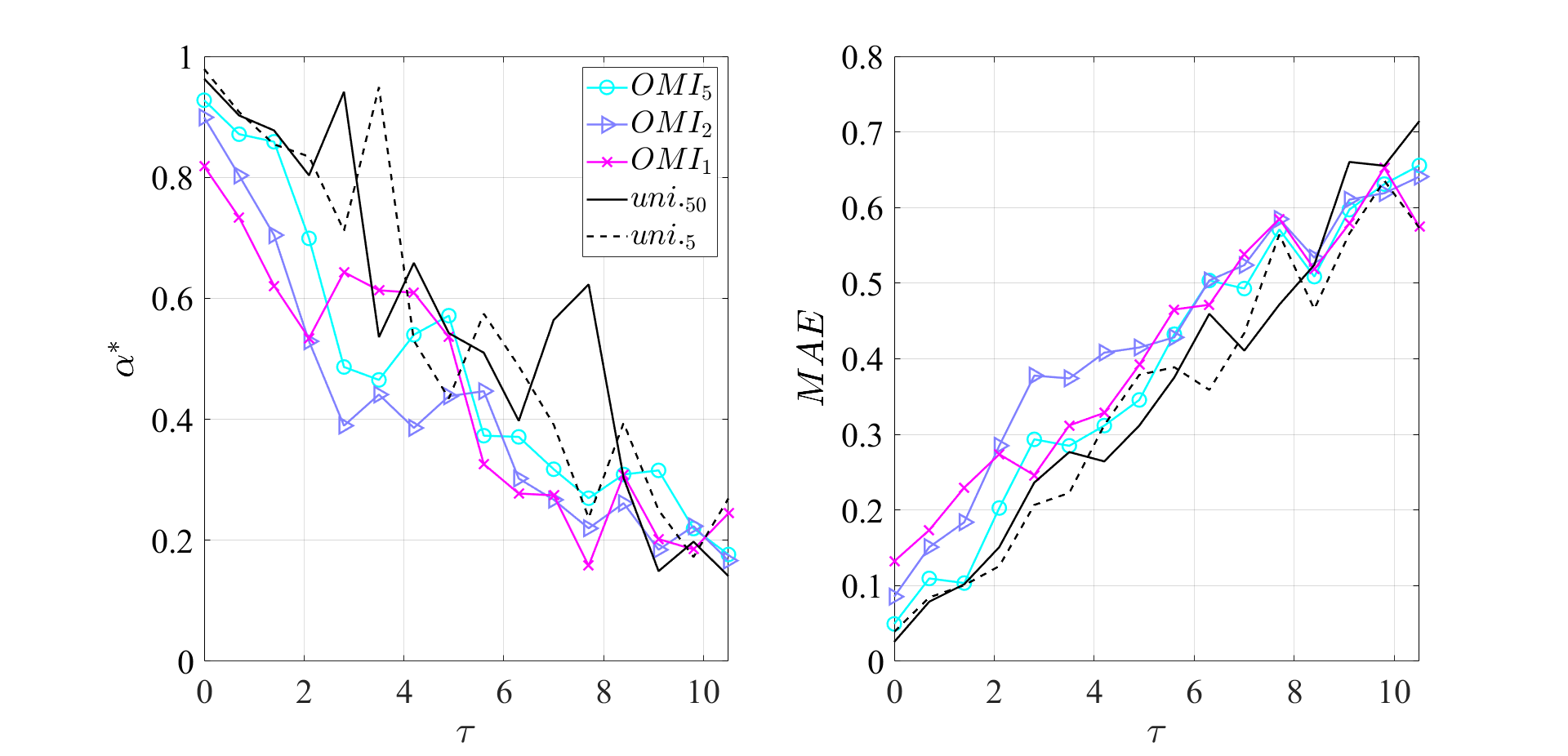}
    \caption{Maximum adjusted area under the precision recall curve, $\alpha^*$ and mean absolute error, $MAE$, for the models summarized in table \ref{tab:NN_models} as a function of the lead time $\tau$.}
    \label{fig:NN_alpha_mae}
\end{figure}

\begin{figure}
    \centering
    \includegraphics[trim = 80 0 0 0, scale = 0.33]{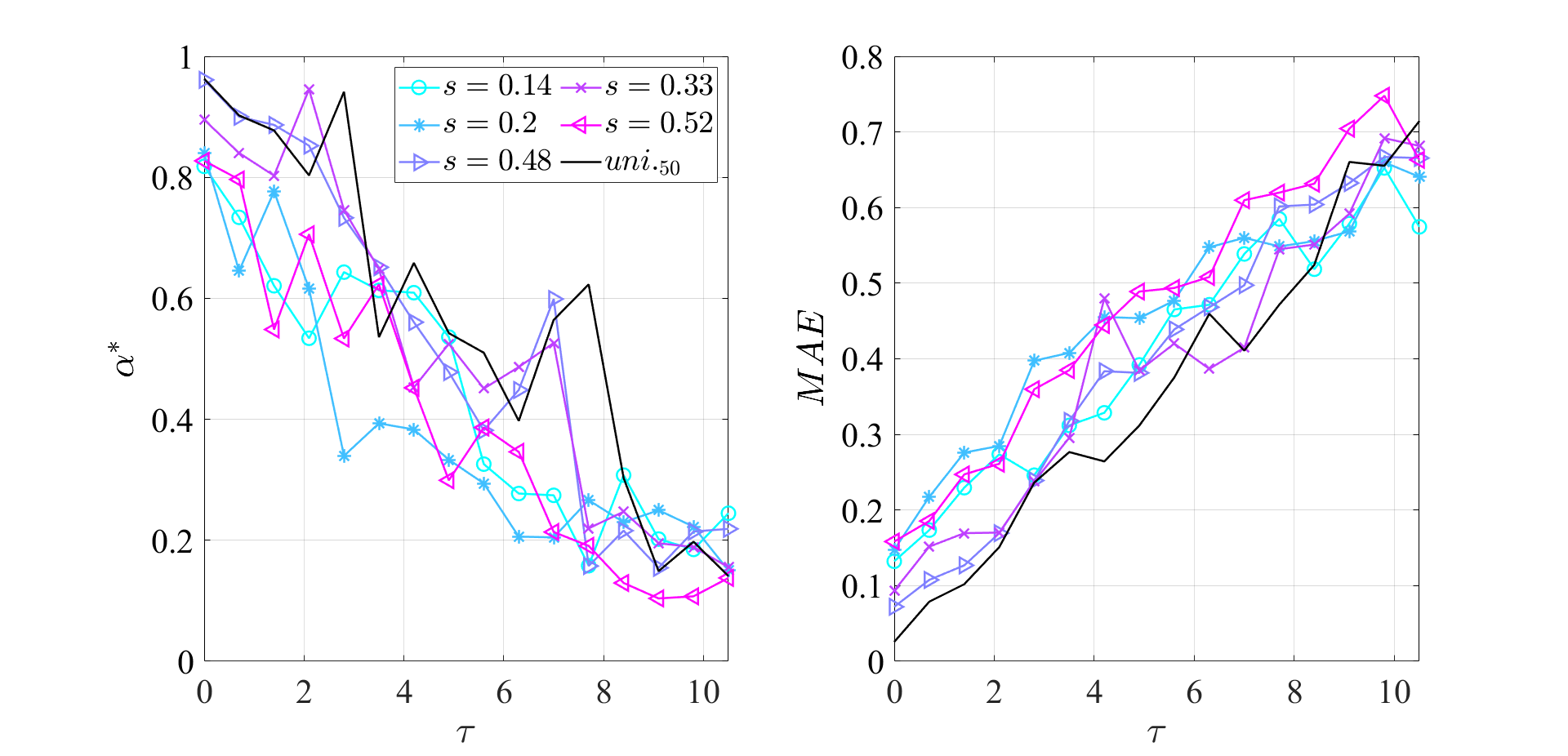}
    \caption{Maximum adjusted area under the precision recall curve, $\alpha^*$ and mean absolute error, $MAE$, for models trained using a single sensor as a function of the lead time $\tau$.}
    \label{fig:NN_alpha_mae_single_sensor}
\end{figure}

Figure \ref{fig:NN_alpha_mae} compares the mean absolute error, MAE, and  $\alpha^*$ of the various models for a range of lead times $\tau$. As expected, for all cases MAE increases and $\alpha^*$ decreases with $\tau$ -- it is more difficult to predict the far future. Comparing the models, we first observe that there is no significant difference between the two reference cases with 5 and 50 sensors, suggesting that the NN model developed by \citet{rudy_prediction_2022}, is amenable to far more sparse sensor distributions than is suggested by those authors. Additionally,  we see no clear distinction between the results of the model trained using our predicted optimal sensor locations and the uniformly sampled reference cases. This highlights the limitations of the mutual information as a practical tool for engineering design. To further highlight this limitation and exclude the possibility of our conclusions being influenced by oversampling we train the model using the 5 optimal sensor locations individually. In other words we train 5 models, each with a single sensor (the $n^{th}$ optimal OMI prediction) as its input. The same error metrics for this experiment are plotted in figure \ref{fig:NN_alpha_mae_single_sensor}. Again we see that the optimal sensor location $s=0.14$ performs no better, and in many cases worse, than the suboptimal locations -- see for example the value of $\alpha^*$ at $\tau = 0$ and $7$. These results strongly suggest that optimal sensing based purely on mutual information does not adequately capture the extreme event mechanism identified in \S\ref{sec:mechanisms} and thus fails as a practical tool for optimal sensing.


\begin{table}[]
    \centering
    \begin{tabular}{c|c|c|c}
        Name & Sensor Placement & Input Observable & Number of Sensors $(N)$  \\
         \hline
        $OMI_N$ & $OMI$ algorithm & $P(t)$ & $N$  \\
        $uni_5$ & $uniform$ & $P(t)$ & $5$  \\
        $uni._{50}$ & $uniform$ & $P(t)$ & $50$  \\
    \end{tabular}
    \caption{Summary of neural net models compared in figure \ref{fig:NN_alpha_mae}.}
    \label{tab:NN_models}
\end{table} 



\section{Wavelet preprocessing for extreme event prediction}\label{sec:wavelet}
Despite its robustness to sensor location, the LSTM network considered in \S\ref{sec:offline} is expensive to train, so here we explore a different avenue of model reduction: preprocessing the data through offline identification of the extreme event dynamics. In \S\ref{sec:mechanisms} we find that the extreme event dynamics are directly related to the dynamics of a single frequency component.  Here we show that exploiting this observation through the event indicator (\ref{gamma}) allows for the forecasting of the extreme events using very simple network models. 
\subsection{Methods}
\begin{figure}
    \centering
    \includegraphics[trim = 0 0 0 0,scale = 0.50]{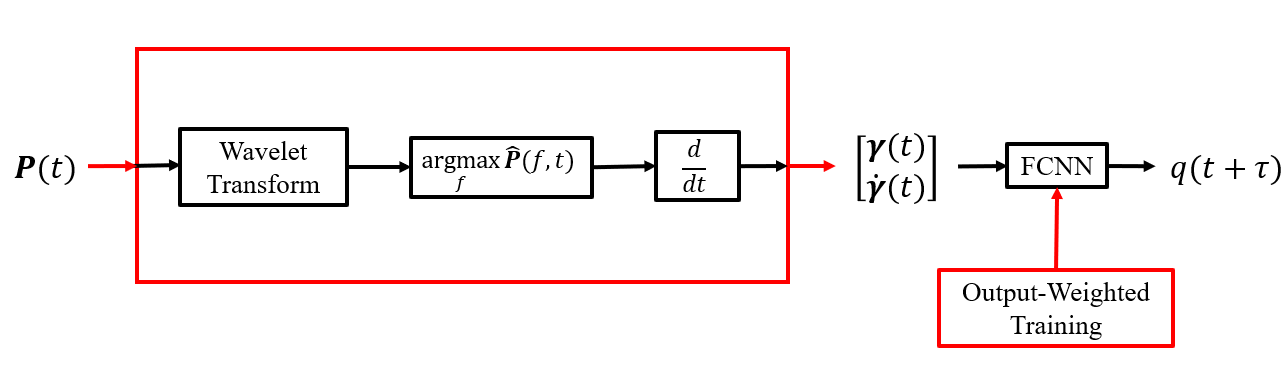}
    \caption{Illustration of wavelet preprocessing algorithm \ref{alg:a2}. Elements outlined in red represent the contributions of this work.}
    \label{fig:flow_chart}
\end{figure}

The extreme event indicator $\boldsymbol{\gamma}$ defined in (\ref{gamma}) is not only highly correlated with the bursting events, as it represents an isolated frequency component, but is also free of noise. This makes it amenable to accurate numerical differentiation. We therefore define the following transformation
\begin{equation}\label{Dn}
    D_{n}\left(\boldsymbol{\gamma}\right): ~\ \boldsymbol{\gamma} \rightarrow \left[ \boldsymbol{\gamma},\dot{ \boldsymbol{\gamma}}, ... , \boldsymbol{\gamma}^{(n)} \right],
\end{equation}
which allows us to track not only the value of $\boldsymbol{\gamma}$, but also its growth rate. This is crucial as we seek to forecast bursting for nonzero lead times $\tau$, and therefore it is imperative for the model to observe growth and not just magnitude. The differentiation operation in (\ref{Dn}) is essentially a phase shift of the signal and thus aids the forecasting capabilities of the model, i.e. the predictions for $\tau >0$. We find that for this flow, a single derivative ($n=1$) is sufficient, and including a second derivative did not meaningfully improve results. Note that this differentiation is applied offline, and thus does not affect the computational cost of training the model. We propose a preprocessing procedure described in algorithm \ref{alg:a2}, in which we replace (\ref{map}) with
\begin{equation}
    \boldsymbol{\Gamma}(t) \rightarrow q(t+\tau).
\end{equation}
Here the input data is defined as
\begin{equation}
    \boldsymbol{\Gamma} \equiv D_{1}\left(\boldsymbol{\gamma}\right) = [\boldsymbol{\gamma},\dot{\boldsymbol{\gamma}}],
\end{equation}
and $\boldsymbol{\gamma}$ is defined in (\ref{gamma}). We utilize a fully connected neural network $f: \mathrm{R}^n \rightarrow \mathrm{R}^1$ with layers
\begin{equation}\label{NN_wavelet}
      \rightarrow FC8 \rightarrow FC16 \rightarrow FC16 \rightarrow FC8 \rightarrow FC1 \rightarrow,
\end{equation}
with the swish activation function \citep{ramachandran_searching_2017} applied between each layer. A regularization constant of $0.01$ was applied to the activation layers -- no regularization was applied to the kernels. Note that unlike the LSTM network (\ref{rudy_net}) utilized by \citet{rudy_prediction_2022}, this network does not map sequences to sequences, it simply maps values of $\boldsymbol{\Gamma}$ at time $t$, to values of $q$ at time $t+\tau$. To train the model we use both a standard and output-weighted mean absolute error loss function,
\begin{equation}
    MAE = \sum_j|\hat{q}_{j} - q_j| ,
\end{equation}
\begin{equation}
    MAE_{OW} = \sum_j\frac{|\hat{q}_{j} - q_j| }{f_q(q_j)}.
\end{equation}
Here $\hat{q}_j$ is the model prediction, $q_j$ is the training data, and $f_{q}(q_j)$ is the probability density function of the training data evaluated at $q_j$.  \citet{rudy_output-weighted_2021}  found that output-weighted loss functions significantly improve prediction of outlier events in a variety of flows including airfoil and Kolmogorov flow. While those authors use the mean square error, we find that in our case the mean absolute error consistently performed slightly better. Our model is summarized graphically in figure \ref{fig:flow_chart}.

In order to quantify the uncertainty of our model, we perform an ensemble analysis resulting in a mean prediction $\hat{q}(t)$ and variance $\sigma_{q}(t)$. We set aside $80\%$ of the data set for training, and for each iteration of the ensemble we randomly select $75\%$ of that training data ($60\%$ of the total) to use for training. The remaining $20\%$ of the data is used for testing. All results presented here are computed exclusively using this test data. 

\begin{algorithm}
\caption{Wavelet Preprocessing}\label{alg:a2}

1. Input data: $\mathbf{P}(t)$ 

2. Compute spectrogram: $|\hat{\mathbf{P}}(t)|$ 

3. Select Maximum: $\boldsymbol{\gamma}(t) \equiv\underset{f\neq f_{v}}{\max}\left(|\hat{\boldsymbol{P}}(f,t)|\right)$


4. Differentiate: $ \boldsymbol{\Gamma} = [\boldsymbol{\gamma},\dot{\boldsymbol{\gamma}}]$

5. Output: $\boldsymbol{\Gamma}(t)$

\end{algorithm}

\subsection{Results: Basis Comparison}\label{sec:wavelet_basis}
To illustrate the advantages of the wavelet basis and the output-weighted loss function we compare the model predictions using the three basis types: $P$, $\Tilde{P}$, and $[\boldsymbol{\gamma},\dot{\boldsymbol{\gamma}} ]$ and the two loss functions $MAE$ and $MAE_{OW}$.  In all cases we use four evenly spread sensor locations at $s = 0.05, 0.35, 0.65, 0.95$. The first two sensor location represent areas identified in \S\ref{sec:MIstruct} and \S\ref{sec:mechanisms} as predictive of the drag. The latter two locations are chosen as to not neglect the underside of the airfoil. In all cases train an ensemble of 10 models for 200 epochs.

\begin{figure}
    \centering
    \subfloat[]{%
    \includegraphics[trim = 70 0 0 0,scale = 0.3]{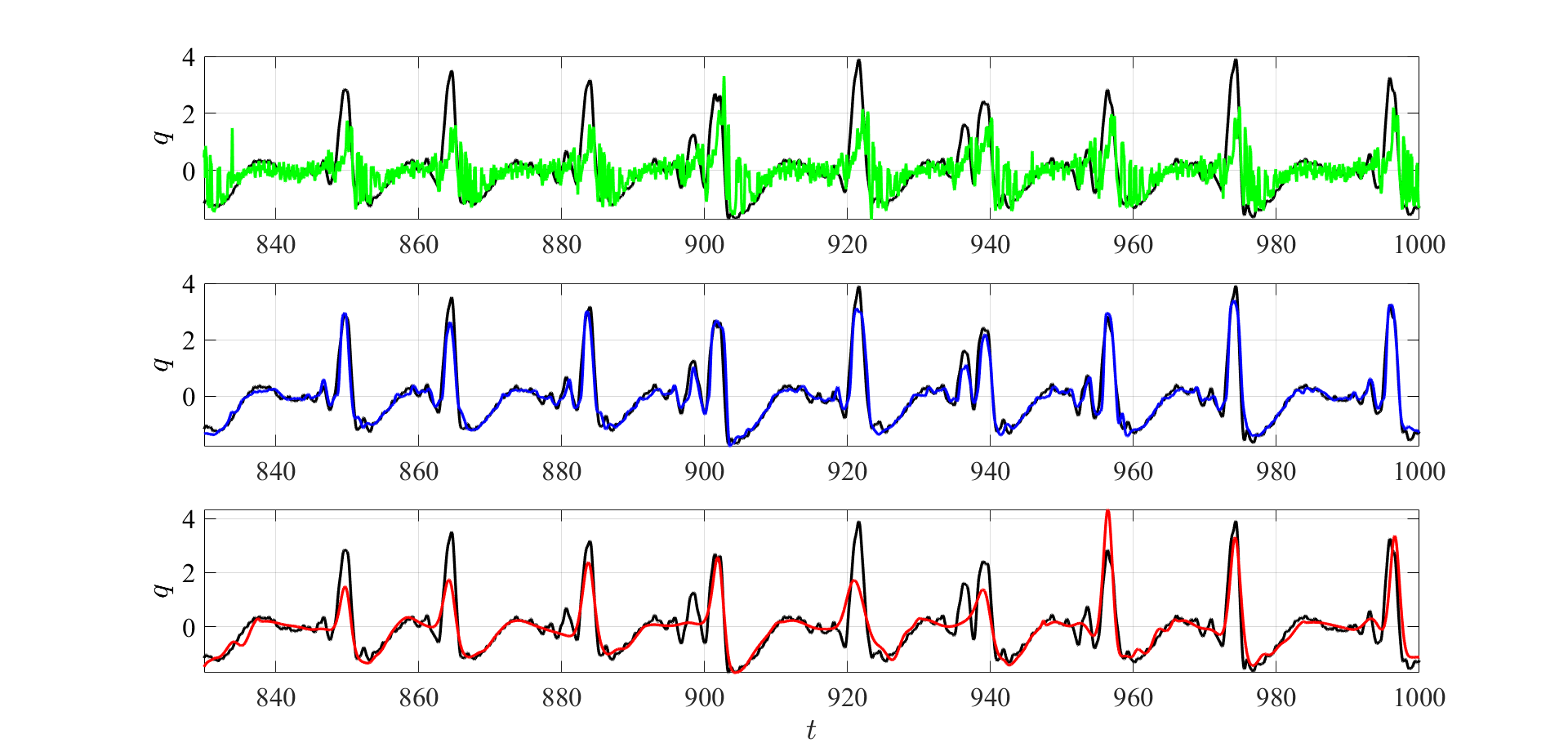}%
    \label{fig:wavelet_net_basis_time_MAE_tau0}
}

\subfloat[]{%
   \includegraphics[trim = 70 0 0 0,scale = 0.3]{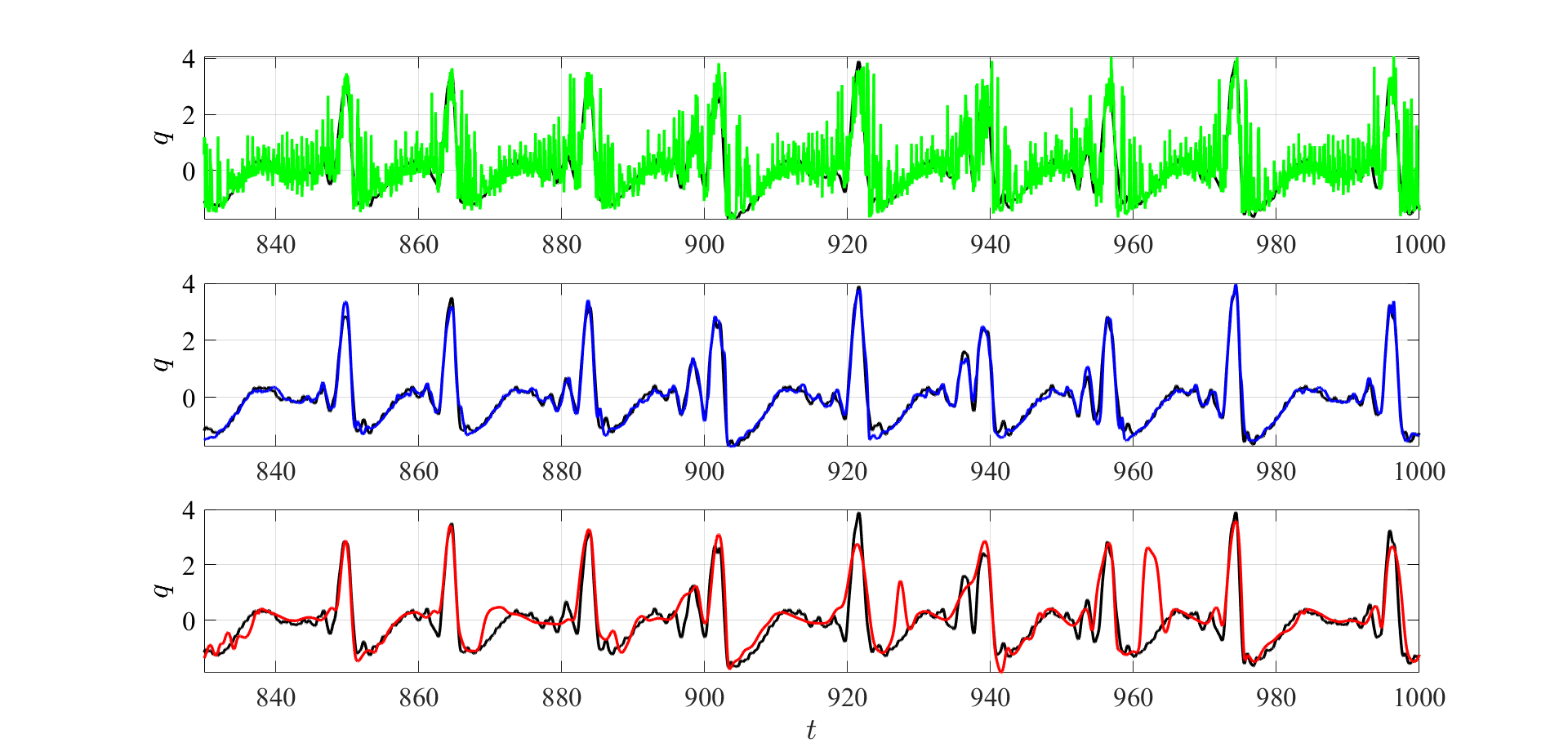}%
    \label{fig:wavelet_net_basis_time_OWMAE_tau0}
}
\caption{Time series of test set predictions for raw pressure (green), filtered pressure (blue), and wavelet (red) models, compared to true values (black) for $\tau = 0$. Loss function: $MAE$ (a), $MAE_{OW}$ (b).}
\label{fig:wavelet_net_basis_tau0}
\end{figure}

\begin{figure}
    \centering
    \subfloat[]{%
    \includegraphics[trim = 70 0 0 0,scale = 0.3]{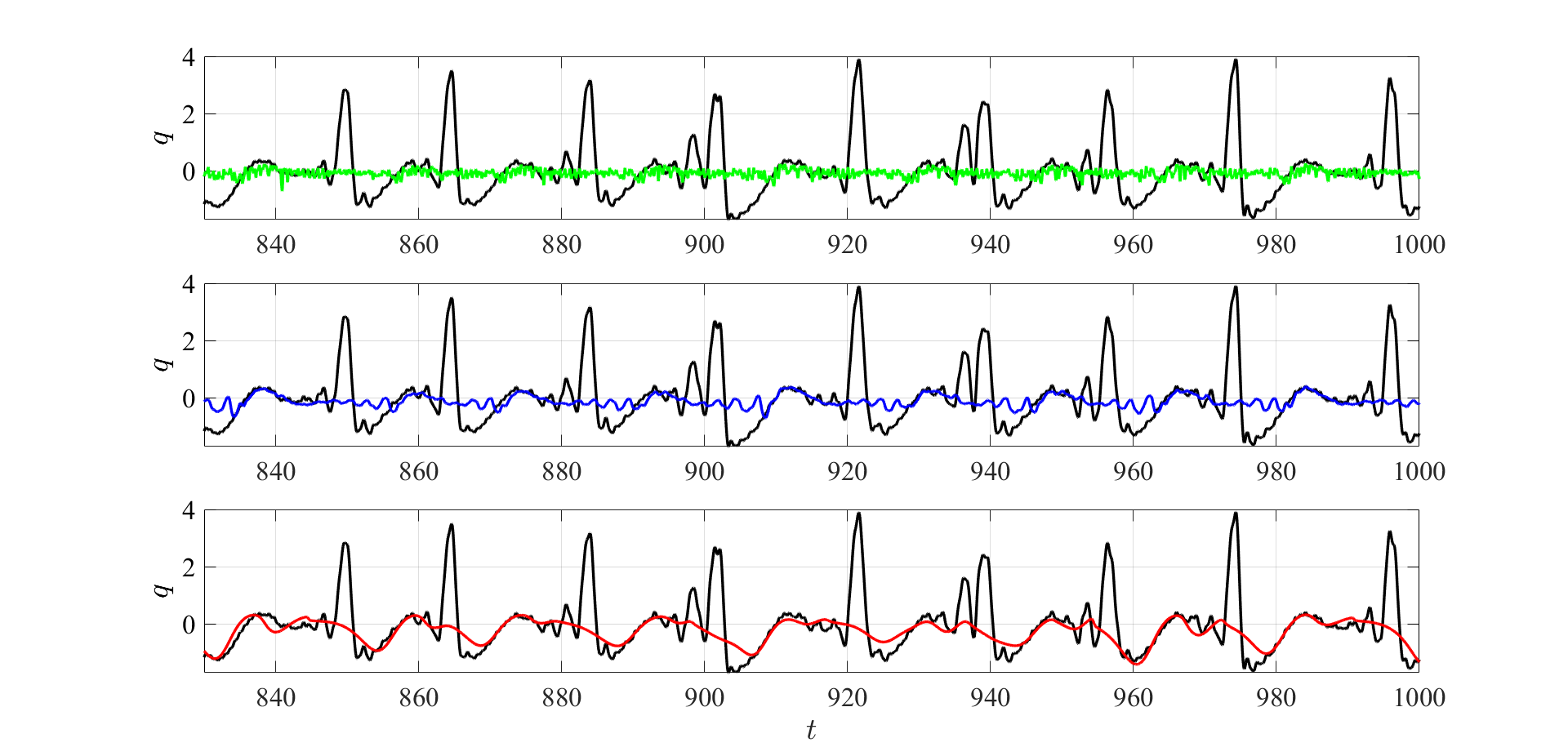}%
    \label{fig:wavelet_net_basis_time_MAE_tau7}
}

\subfloat[]{%
   \includegraphics[trim = 70 0 0 0,scale = 0.3]{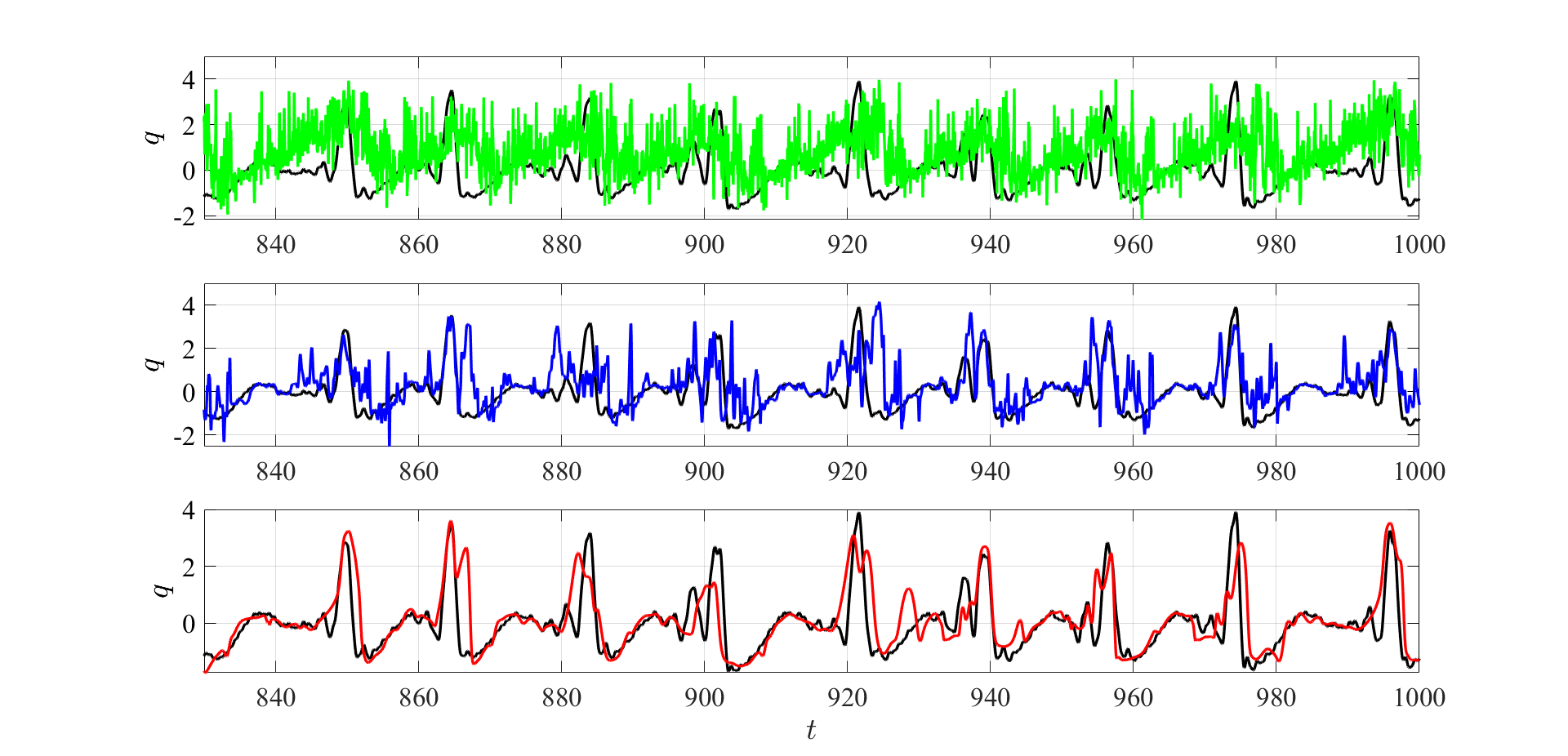}%
    \label{fig:wavelet_net_basis_time_OWMAE_tau7}
}
\caption{Time series of test set predictions for raw pressure (green), filtered pressure (blue), and wavelet (red) models, compared to true values (black) for $\tau = 7$. Loss function: $MAE$ (a), $MAE_{OW}$ (b).}
\label{fig:wavelet_net_basis_tau7}
\end{figure}

\begin{figure}
    \centering
    \subfloat[]{%
    \includegraphics[trim = 70 0 0 0,scale = 0.3]{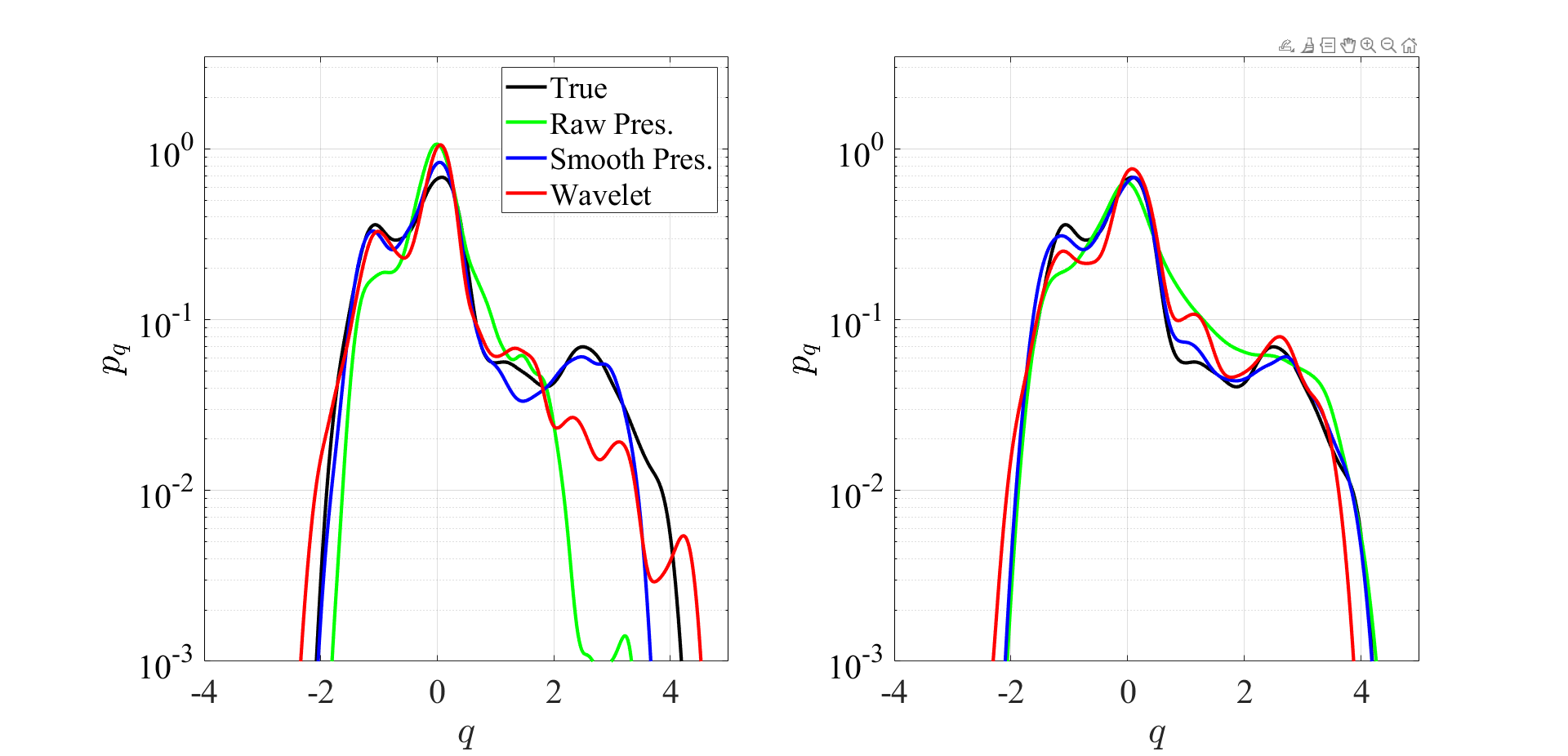}%
    \label{fig:wavelet_net_basis_pdf_tau0}
}

\subfloat[]{%
   \includegraphics[trim = 70 0 0 0,scale = 0.3]{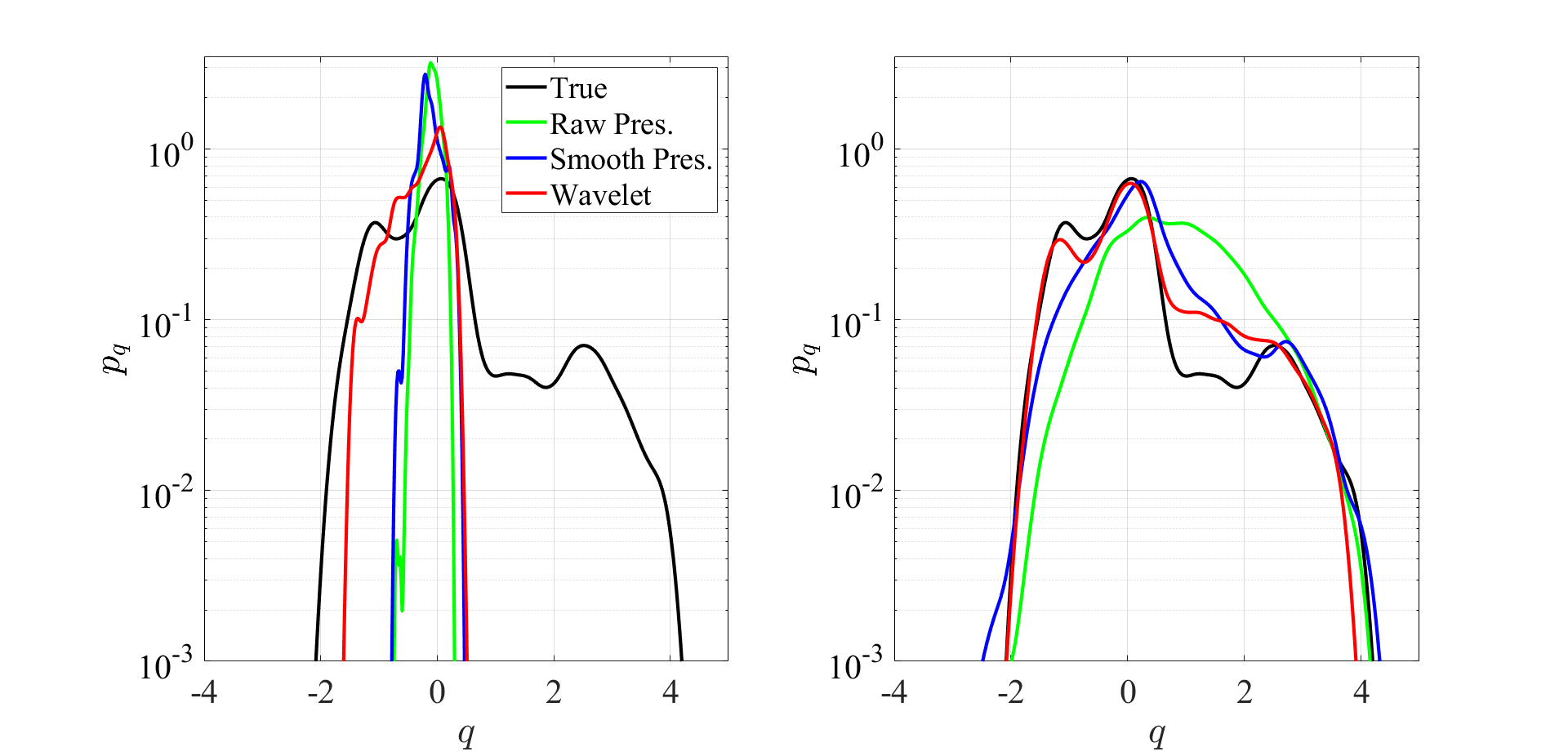}%
}
\caption{Probability density function of test set predictions for raw pressure (green), filtered pressure (blue), and wavelet (red) models, compared to true values (black) for $\tau = 0$ (a) and $\tau =7$(b). Loss function: $MAE$ (left), $MAE_{OW}$ (right). }
\label{fig:wavelet_net_basis_pdf}
\end{figure}

Figures \ref{fig:wavelet_net_basis_tau0}  and \ref{fig:wavelet_net_basis_tau7} compare the model prediction of the drag coefficient to the true value for $\tau = 0$ and $\tau = 7$. At $\tau =0$, the filtered pressured model and the wavelet model, shown in blue and red respectively, perform well, with only the raw pressure model, shown in green, suffering from significant noise corruption. In fact, in this case the smooth pressure model slightly outperforms the wavelet basis -- best seen by comparing the predicted probability density functions in figure \ref{fig:wavelet_net_basis_pdf}. This is because there exists an accurate linear mapping from the smooth pressure $\Tilde{\mathbf{P}}(t)$ to the drag $q(t)$, and in taking the wavelet transform of the filtered pressure some information is lost -- limiting the potential accuracy of the wavelet model. See appendix \ref{app:linear} for a brief discussion on this linear mapping. Furthermore, we observe that for zero lead time, the models trained using the $MAE_{OW}$ loss function perform slightly better than those trained using the $MAE$ loss -- The benefit is most pronounced for the raw pressure data.

The benefits of the wavelet basis and the output weighted loss functions do not become apparent until considering non-zero lead times -- a reflection of the nonlinearity of the time shift operation $q(t) \rightarrow q(t+\tau)$. In this case, when using the standard $MAE$ loss, all three models entirely fail to capture the extreme events. However, using the output-weighted loss the wavelet basis retains much of the accuracy observed for $\tau = 0$, while the performance of the raw and filtered pressure models deteriorate significantly. The filtered pressure model still traces the occurrence of the extreme events, but suffers from significant noise corruption leading to number of false positive predictions. This phenomenon is even more pronounced for the raw pressure model, which, as expected, suffers from even greater noise corruption. The distinction between the three basis types is less evident in the predicted probability density functions shown in \ref{fig:wavelet_net_basis_pdf}. Here we  see again that the models trained using the standard $MAE$ loss fail to capture the tails of the distribution entirely, however with the $MAE_{OW}$ loss both the filtered pressure and wavelet models capture the general shape of the distribution. The wavelet model does however capture the small peaks around $q=-1.5$ and $q=3$ slightly better than the others.

To quantify the forecasting capabilities of each model we track the number of extreme events predicted as a function of time. For this purpose we define an extreme event as a local maximum, whose value is more than 2 standard deviations greater than the mean. Thus, a time instant $t_j$ is considered to represent an extreme event $t_{EE}$ if it satisfies the following conditions,
\begin{equation}\label{EE_def}
    t_{EE}: ~\ t_j   ~\  s.t. ~\ \left.\left[\frac{\partial q}{\partial t}\right|_{t_j} = 0 ~\ \& ~\ q(t_j) > 2\sigma_q \right].
\end{equation}
We then define the number of extreme events $N_{EE}(t_1,t_2)$ as the number of extreme events in the interval, $t_1$ to $t_2$, or more explicitly,
\begin{equation}
    N(t_n,t_0) = \sum_{j=j_0}^{j_n} \delta_{t_j,t_{EE}},
\end{equation}
where $t_j$ is treated as a discrete series and $j_0$ and $j_n$ are the indices of $t_n$ and $t_0$ respectively. In order to avoid over-penalizing noise, we enforce a minimum separation of the identified extreme events equal to the characteristic period of the extreme event frequency: $T_{EE}=1/f_{EE}$. While it relies on two user defined parameters: the extreme event threshold and the minimum peak separation, this metric provides a useful quantification of the forecasting capabilities of each model.

\begin{figure}
    \centering
     \includegraphics[trim = 70 0 0 0,scale = 0.3]{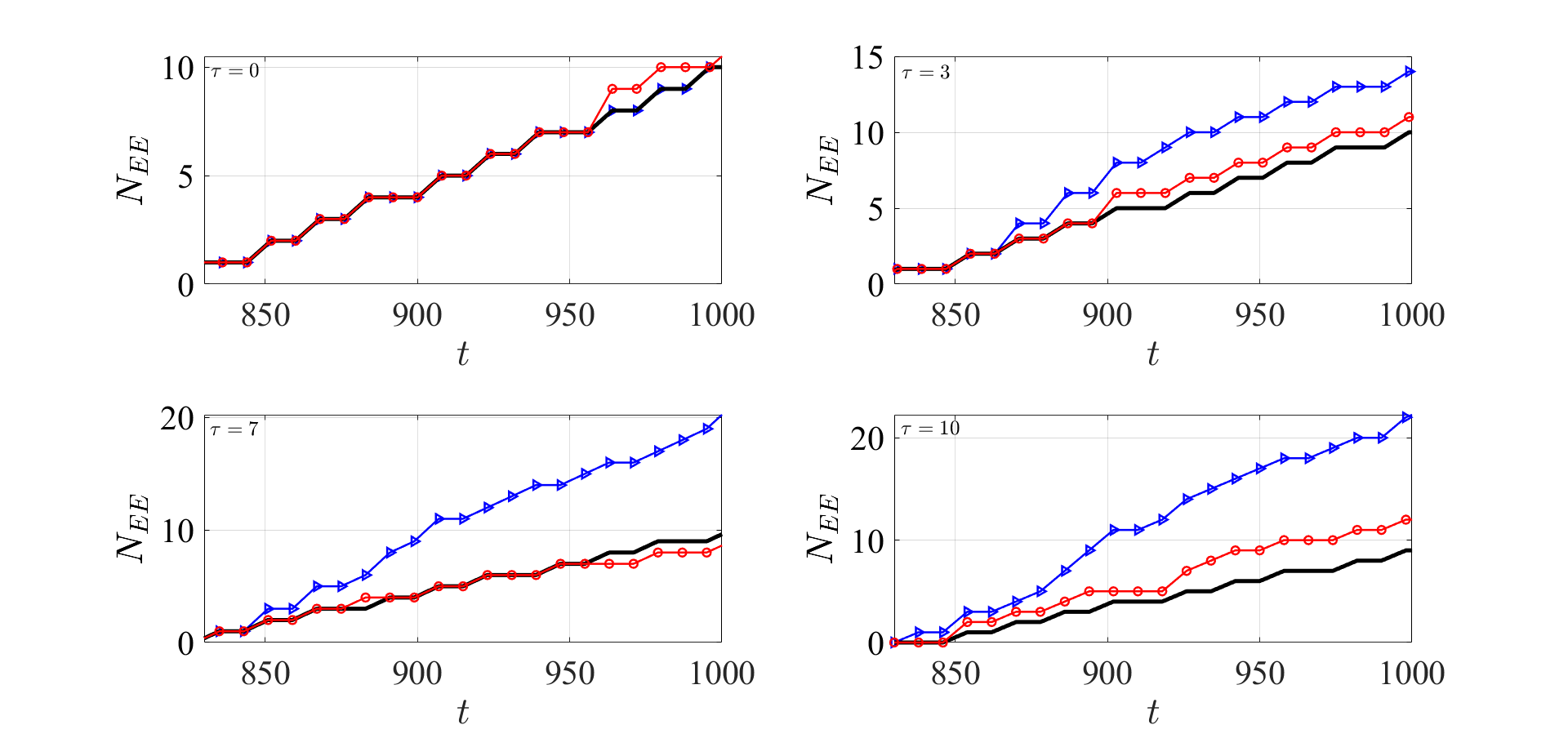}%
    \caption{Predicted number of extreme events by filtered pressure model (blue) and wavelet model (Red) compared to true number (black) for from top left $\tau = [0,3,7,10]$. }
    \label{fig:wacelet_net_basis_EE_metric}
\end{figure}

This metric is plotted in figure \ref{fig:wacelet_net_basis_EE_metric} for the filtered pressure and wavelet models for $\tau = [0,3,7,10]$ -- we omit the raw pressure model due to its poor performance. Since the models trained using the $MAE$ loss and perform so poorly for $\tau >0$ we plot only the results obtained using the $MAE_{OW}$ loss. Again we see that for $\tau = 0$ both models perform similarly, with the filtered pressure model slightly outperforming the wavelet model. However, as the lead time $\tau$ increases, the wavelet model retains much of it's accuracy, while the filtered pressure model on the other hand dramatically overestimates the number of extreme events. This is due to the significant noise in the filtered pressure model, the magnitude of which is often comparable to the underlying signal. 

\begin{figure}
    \centering
     \includegraphics[trim = 70 0 0 0,scale = 0.3]{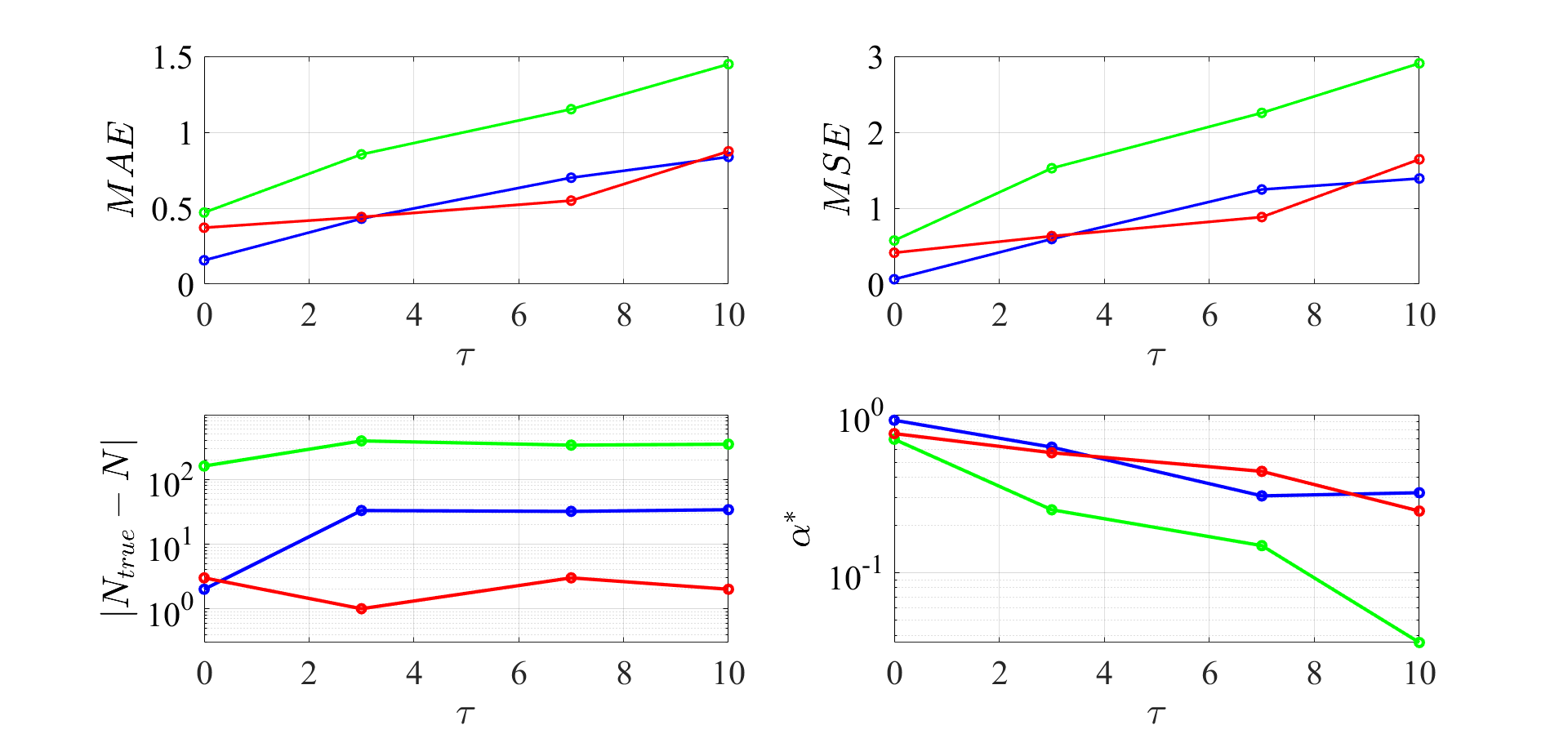}%
    \caption{From top left: $MAE$, $MSE$, error in number of predicted extreme events, and $\alpha^*$. Models:  raw pressure (green), filtered pressure (blue), and wavelet (red).}
    \label{fig:wacelet_net_basis_error}
\end{figure}

We also compute the MAE, MSE, $\alpha^*$, defined in (\ref{guth_crit}), as well as the error in the total number of extreme events predicted. These are plotted in figure \ref{fig:wacelet_net_basis_error}. Again, both pressure models overestimate the number of extreme events defined by (\ref{EE_def}), however for the aggregate error metrics ($MAE$ and $MSE$) as well as the $\alpha^*$ the differences are less pronounced. Both the filtered pressure and wavelet model significantly outperform the raw pressure model, but the difference between them is minimal. 

The discrepancy in performance between the three basis types can be understood through the simple nature of the model architecture. The large amplitude of the high frequency fluctuations in the raw pressure is comparable and sometimes even larger than the bursting amplitude -- see figure \ref{fig:example_data} -- therefore a one-to-one map is destined to fail. This phenomenon is mitigated by filtering the vortex shedding frequency out of the pressure data -- in this case there is indeed a linear map for zero lead time. However, for non-zero lead times the amplitude of the fluctuations at the extreme event frequency are significant enough to introduce significant ambiguity in a one-to-one map. Conversely, the wavelet basis is free of noise and fluctuates on a time scale associated with the mean time between extreme events, thereby greatly improving the feasibility of such a simple mapping. Moving forward we exclusively use the wavelet pre-processed input data $\boldsymbol{\Gamma}$.



\subsection{Results: Optimal Sensing}\label{sec:wavelet_search}
We now assess how to best exploit the wavelet preprocessing algorithm through optimal sensor selection. Here we focus exclusively on $\tau = 7$. As in \S\ref{sec:offline} we use algorithm \ref{alg:a1}, however in this case the acquisition function requires training the network (and is minimized, not maximized). In particular we consider the following two acquisition functions which penalize uncertainty in the model prediction,
\begin{equation}\label{iu}
    a_{iu} = \frac{1}{T}\int^{T}_0 \sigma_q(t) dt = \frac{1}{N}\sum^{N}_{j=1} \sigma_{q,j},
\end{equation}

\begin{equation}\label{pw}
    a_{pw} = \frac{1}{T}\int^{T}_0 \frac{\sigma_q(t)}{p_q(q(t)) }dt = \frac{1}{N}\sum^{N}_{j=1} \frac{\sigma_{q,j}}{p_q(q_j)}.
\end{equation}
We refer to these as \textit{``integrated uncertainty'' (iu)} and \textit{``probability weighted'' (pw)} respectively. To distinguish the effects of the loss and acquisition functions, we perform three iterations of algorithm \ref{alg:a1} with each acquisition function with both the $MAE$ and $MAE_{OW}$ loss functions. As the acquisition functions need to be evaluated at every sensor location, we use a slightly reduced ensemble of 7 models and train over only 70 epochs during the active search. The resulting optimal sensor locations are summarized in table \ref{tab:wavelet_sensing}. Once the optimal sensor locations are found, we retrain an ensemble of 10 models using those optimal sensors locations for 200 epochs.

\begin{figure}
    \centering
    \subfloat[]{%
    \includegraphics[trim = 70 0 0 0,scale = 0.3]{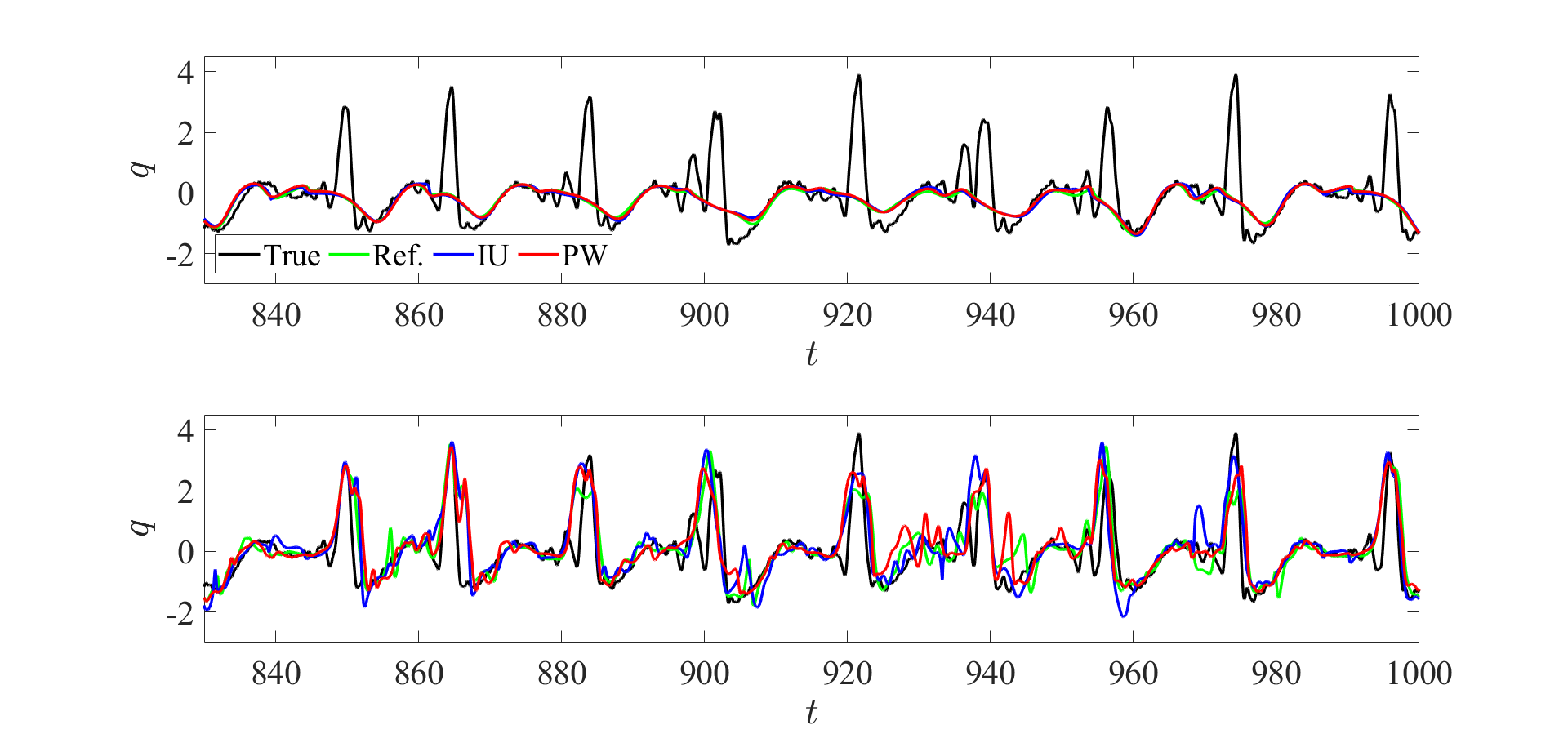}%
    \label{fig:wavelet_net_aqFun_time}
}

\subfloat[]{%
   \includegraphics[trim = 70 0 0 0,scale = 0.3]{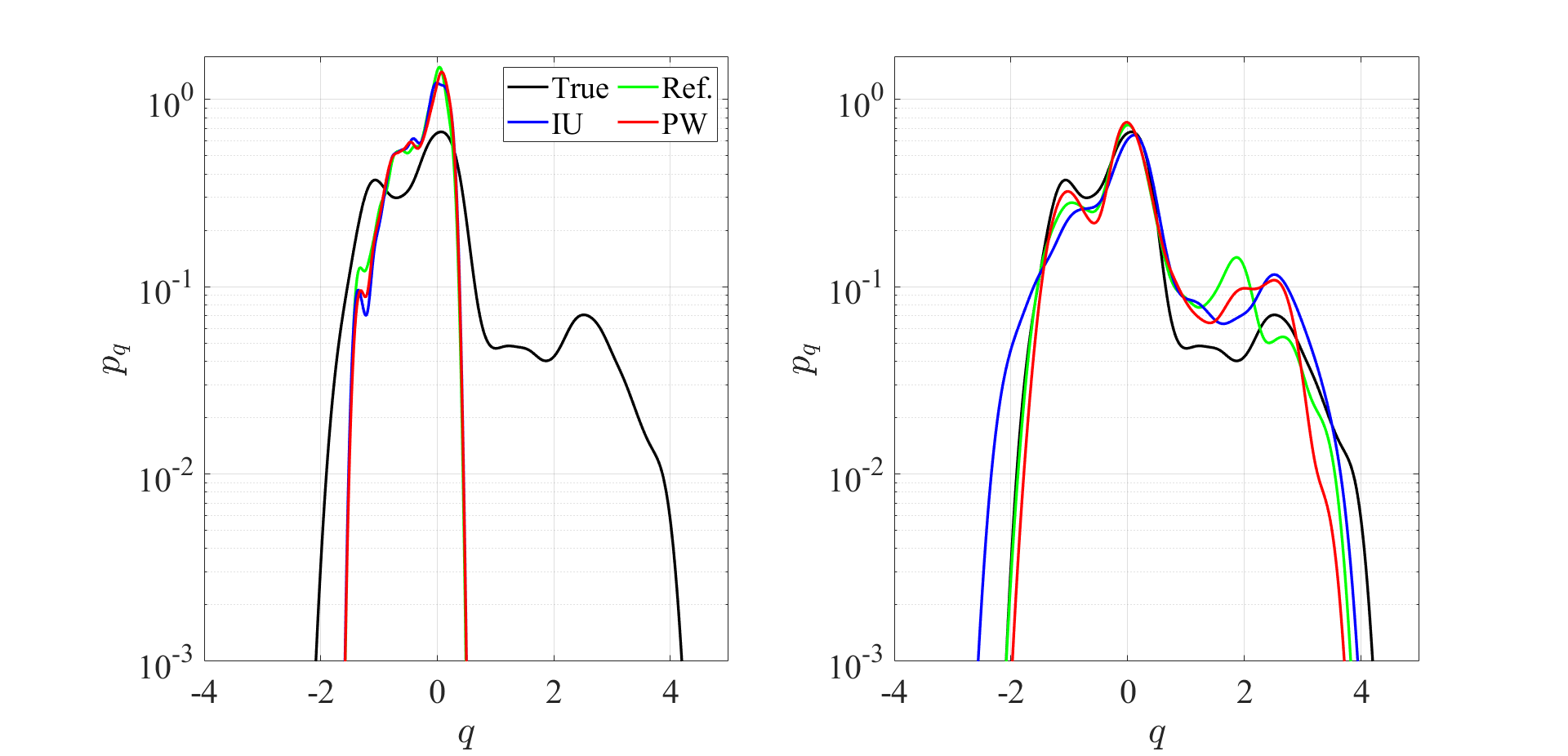}%
    \label{fig:wavelet_net_aqFun_pdf}
}

\caption{Comparison of test set predictions for reference (green), $IU$ (blue), and $PW$ (red) acquisition functions to true values (black). Time series for $MAE$ (top) and $MAE_{OW}$ (bottom) (a), and probability density function for $MAE$ (left) and $MAE_{OW}$ (right) (b).}
\label{fig:wavelet_net_aqFun}
\end{figure}

\begin{figure}
    \centering
    \includegraphics[trim = 90 0 0 0,scale = 0.33]{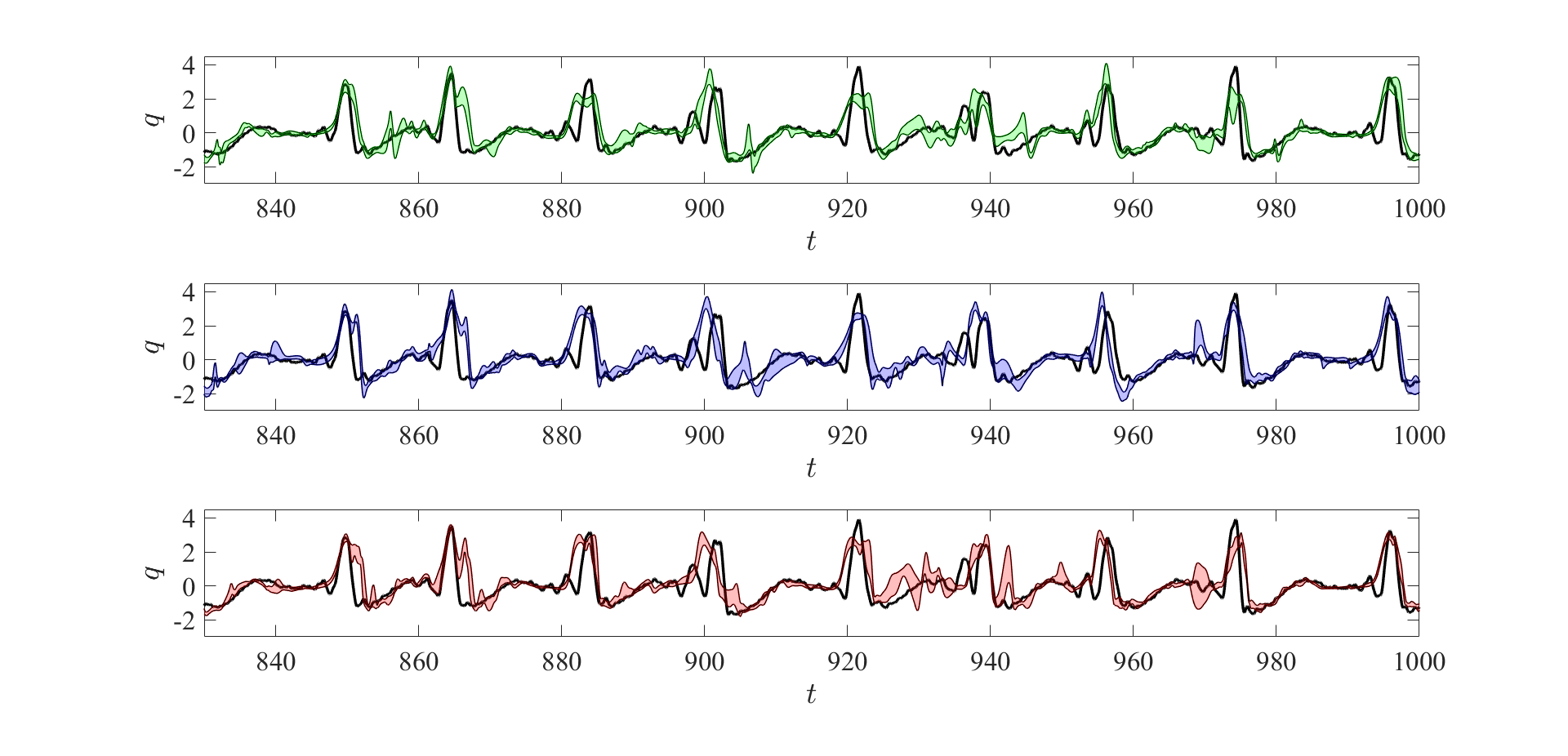}
    \caption{Comparison of test set predictions for reference (green), $IU$ (blue), and $PW$ (red) acquisition functions to true values (black). Shaded area represents mean $\pm$ one standard deviation of ensemble prediction.}
    \label{fig:wavelet_net_aqFun_time_var}
\end{figure}

Figure \ref{fig:wavelet_net_aqFun} compares the  output pdf and time series of the mean predictions  of the model trained with each loss function and each acquisition function -- each using their respective optimal sensor locations.  As a comparison we also include the predictions of a reference model with three evenly spaced sensors. For both the time series and the pdf, we observe that regardless of the acquisition function the models trained using the standard $MAE$ loss fail, while the models trained with the output-weighted $MAE_{OW}$ predict the bursting events relatively accurately. To further compare the models we plot the time series of the uncertainty bound, $\hat{q}(t) \pm \sigma_{q}$ in figure \ref{fig:wavelet_net_aqFun_time_var}. Consistent with the results of \S\ref{sec:offline}, we observe the model to be robust to specific sensor locations, with little distinction between the three sensor distributions.

Interestingly, inspection of figures \ref{fig:wavelet_net_aqFun_time} and \ref{fig:wavelet_net_aqFun_time_var} indicates that the output-weighted acquisition function (\ref{pw}) performs slightly worse than the others -- exhibiting some false positive fluctuations between $t=920$ and $t=960$. Furthermore, the results using the non output-weighted acquisition function (\ref{iu}) do not exhibit any meaningful improvement over the reference case. These findings indicate that the specific locations of the sensors are of secondary importance when compared to the effects of the output-weighted loss function and the wavelet preprocessing. The latter of which extracts the bursting events from the input data a priori. These results indicate that further emphasizing extreme events through strategies such as output-weighted optimal sensing is not only unnecessary, but could result in a loss of accuracy during the quiescent periods.

From a practical point of view, the similarity of these results is significant. The active regions of the flow, and thus the optimal sensor locations predicted here and in \S\ref{sec:offline}, are likely to vary with Reynolds number and angle of attack. However, aircraft experience a wide range of flow speeds and orientations, making the robustness to sensor location a valuable asset. These results support the possibility of a sparse sensing strategy which is applicable for a wide range of airfoil designs and is robust to dynamic changes in angle of attack. A parameter study over Reynolds number and flow geometry to confirm this hypothesis is beyond the scope of the present work, but is the topic of ongoing research.

\begin{table}[]
    \centering
    \begin{tabular}{c|c|c}
        & $IU$ & $PW$ \\
        \hline
        $MAE$ & $[0.31,0.15,0.25] $& $[0.09,0.84,0.05]$  \\
        $MAE_{OW}$ & $[0.21,0.15,0.60] $& $[0.58,0.76,0.13]$
    \end{tabular}
    \caption{Optimal sensor locations predicted using algorithm \ref{alg:a1} with acquisition functions (\ref{iu}) and (\ref{pw}) with $MAE$ and $MAE_{OW}$ loss functions.}
    \label{tab:wavelet_sensing}
\end{table}

\FloatBarrier

\newpage
\section{Discussion}\label{sec:discussion}
We have investigated the mechanisms driving the non-periodic bursting phenomena observed in the two-dimensional flow over a NACA 4412 airfoil at finite angle of attack. We have conducted a detailed analysis of the spatiotemporal statistics of the airfoil surface pressure and its connection to the extreme events observed in the drag force. Through a wavelet analysis we found that the surface pressure exhibits multi-scale behaviour with three distinct time scales. In addition to the dominant vortex shedding frequency, the flow exhibits a slowly varying quiescent time scale and a second energetic frequency component -- at approximately one third the vortex shedding frequency. We established that the extreme excursions of the drag first observed by \citet{rudy_prediction_2022}, correspond to instabilities of this latter frequency component in the surface pressure.

These findings were corroborated by an analysis of the wavelet transformed vorticity field. This analysis revealed that during quiescent times the extreme event manifold evolves independently of the vortex shedding manifold, however occasionally the extreme event manifold undergoes a transient instability which links the fortunes of these two generally disparate time scales. This instability is comprised of two steps, first the extreme event manifold draws energy from the higher frequency vortex shedding flow, then at the extreme event frequency, there is an abrupt nonlinear energy transfer from smaller to larger spatial length scales. Interestingly these findings are contrary to the far more common situation where linear instabilities transfer energy from a slowly evolving (or stationary) mean flow to faster time scales and smaller length scales. Therefore, while we have identified the slow-fast system at the heart of the bursting events, the exact mechanism by which the instabilities in the pressure and vorticity are translated to the aerodynamic forces is not yet clear, and remains the topic of ongoing research.
For example, it is still unclear what causes the global (integrated over the full domain) magnitude of the extreme event mode to decrease during the extreme events -- see the lower panel of figure \ref{fig:wavelet_norms} -- or why the temporal correlation of the vortex shedding mode exhibits fluctuations resembling the extreme event frequency -- see the upper panel of the same figure. Furthermore, here we have considered only a single angle of attack, and further study is required to establish how the orientation of the flow impacts both the active regions of the airfoil and the extreme event frequency.

From a modeling perspective, we pursued two separate strategies. First, in \S\ref{sec:offline} we investigated the implications of these results for the existing LSTM architecture developed by \citet{rudy_prediction_2022} -- which takes raw pressure as its input. We considered an optimal sensing strategy based purely on the statistics of the data, and therefore did not require the computationally costly step of training model. Using the LSTM model this mutual information based algorithm failed to predict sensor locations which performed better than a simple uniform sensor distribution. This failure of the purely mutual information based sensor placement demonstrated the limitations of a purely statistical offline sampling strategy and highlighted the limitations of mutual information as a practical tool. Additionally, the model complexity incurred by the LSTM layers needed to process rapidly varying time series such as the fluctuating surface pressure remains cumbersome regardless of the sparsity of the sensor array.

Second, we also developed a preprocessing algorithm (see figure \ref{fig:flow_chart}) to extract the time varying magnitude of the extreme event frequency component from the raw pressure signal. By isolating the wavelet coefficient associate with the extreme event frequency we eliminate the high frequency fluctuations resulting in a signal which slowly fluctuates on a time scale associated with the mean time between extreme events -- which are by definition rare. This enables the (approximate) one-to-one mapping of the wavelet coefficient, which is free of rapid high amplitude noise, to the drag for lead times $\tau >0$. This then eliminates the need for a costly LSTM architecture and allows for accurate prediction using a simply connected feed forward neural network. These results are consistent with findings of \citet{cousins_sapsis,cousins_reduced-order_2016} who used a spacial wavelet transform wavelet transform to extract unstable spatial length scales to efficiently predict rogue waves in variety of dynamical systems including the Majda–McLaughlin–Tabak model and the modified nonlinear Schrodinger equation.

While this preprocessing drastically reduces the noise in the signal, it can, and in our case does, eliminate some potentially useful information as well. As noted in \S\ref{sec:wavelet_basis} and in appendix \ref{app:linear} for $\tau = 0$ there exists an accurate linear map from the surface pressure to the drag coefficient. By isolating a single wavelet coefficient, some of the information in the pressure signal is lost leading to the wavelet model performing slightly worse than the filtered pressure model for $\tau =0$. However, the predictions from the wavelet model are far more robust to increasing values of $\tau$. At $\tau = 7$ the predictions of the wavelet model have degraded only slightly, while the raw and filtered pressure models exhibit significant noise corruption. The higher the frequency of oscillation, the more nonlinear the transformation $q(t) \rightarrow q(t+\tau)$ -- resulting in the degradation of the filtered and raw pressure models for $\tau >0$.

This preprocessing alleviates the need for recursive or convolutional network architectures as used by authors such as \citet{hou_machine-learning-based_2019,rudy_prediction_2022}. However, even with this highly extreme event targeted algorithm we found that training the model using an output-weighted loss function is necessary for accurate predictions. Most interestingly, we find that with these training interventions the specific locations of the sensors is of secondary importance. This is incredibly advantageous as it suggests that the predictive capabilities of our approach are robust to dynamic changes in angle of attack or free stream velocity -- however this requires further study. Our findings suggest that improving the prediction of rare events does not necessarily require more complex models, but can be achieved by identifying observables which reflect the underlying physical mechanisms and through tailored training strategies -- as also discussed by various authors including \citet{farazmand_variational_2017,sapsis_output-weighted_2020,rudy_output-weighted_2021,blanchard_output-weighted_2021}. We believe the herein proposed wavelet based analysis is applicable to a wide range of slow-fast systems and remains the topic of ongoing research.







\newpage
\appendix
\section{Appendix}

\subsection{Linear Mapping}\label{app:linear}
\begin{figure}
    \centering
    \includegraphics[trim = 70 0 0 0,scale = 0.3]{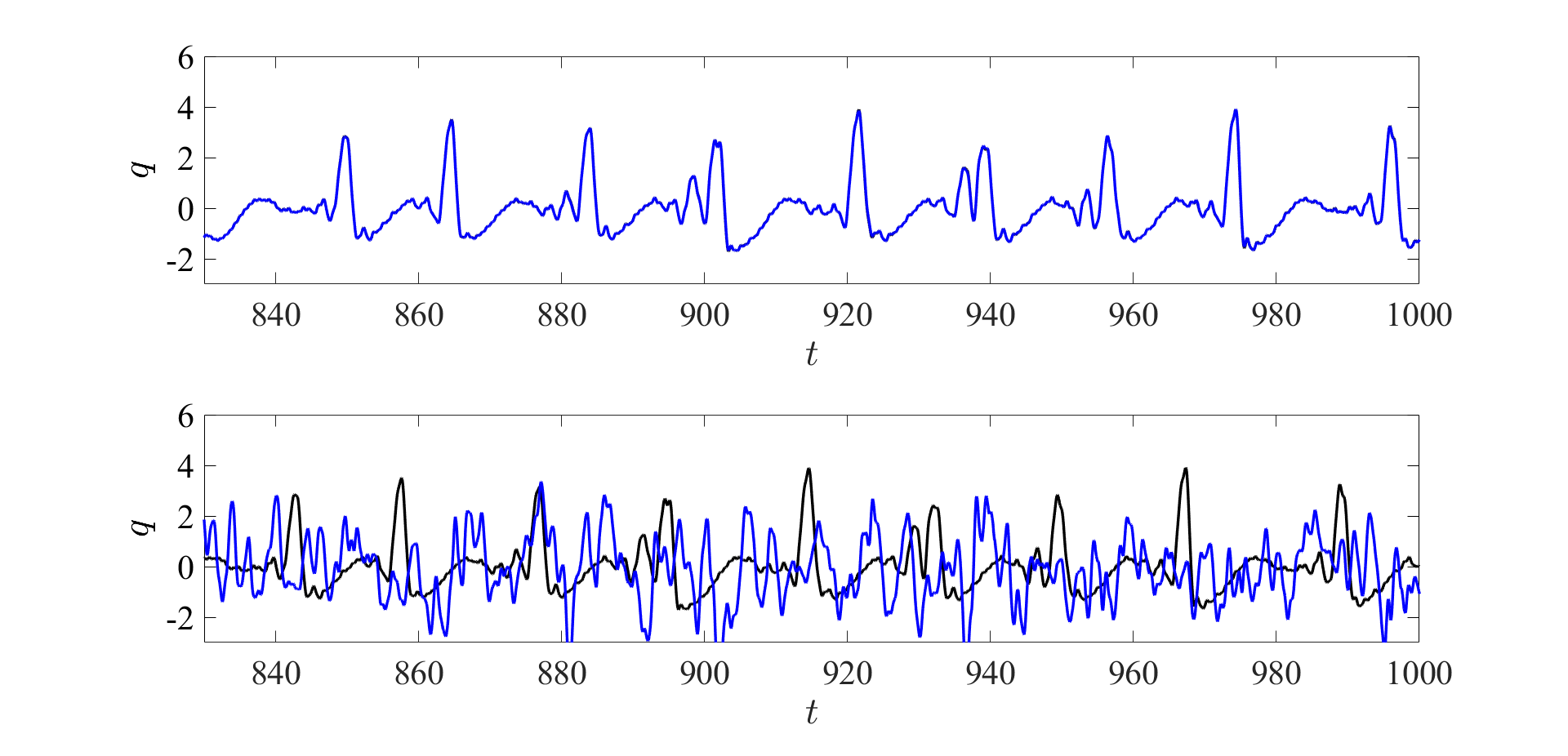}
    \caption{Prediction of drag coefficient from  linear regression model (blue) compared to true value (black). $\tau = 0$ (upper panel), $\tau = 7$ (lower panel).}
    \label{fig:linear}
\end{figure}

Here we illustrate the linear map from the vector valued filtered pressure signal $\Tilde{\mathbf{P}}(t)$ to the scalar drag coefficient $q(t)$. Let $\Tilde{\mathbf{P}} \in \mathbb{R}^{N\times 100}$ and $\mathbf{q} \in \mathbb{R}^{N\times 1} $ be their discrete representations -- $N$ is the number of data points (time steps). We then seek a linear representation $\hat{\mathbf{q}} = \Tilde{\mathbf{P}}\mathbf{a}$ that minimizes the $L_2$ norm $\|\mathbf{q}-\hat{\mathbf{q}}\|^2$. The optimal coefficient vector is given by
\begin{equation}\label{lin_regresssion}
    \mathbf{a}^* = \Tilde{\mathbf{P}}_{train}^+\mathbf{q}_{train},
\end{equation}
where $^+$ denotes the pseudo inverse, and the subscript $_{train}$ refers to the subset of data used for training. Here we use the first $10\%$ of the data to fit the regression (\ref{lin_regresssion}) and the last $20\%$ for testing. Figure \ref{fig:linear} compares the predictions of the linear regression to the truth for $\tau =0$ and $\tau = 7$. For $\tau = 0$ the linear prediction is indistinguishable from the truth, while for $\tau = 7$ the linear model completely fails. This is a reflection of the highly nonlinear nature of the time shift operation $q(t) \rightarrow q(t+\tau)$.

\FloatBarrier
\section*{Funding Sources}

\section*{Acknowledgments}
We thank Samuel Rudy and Tanner Harms for their constructive feedback. We also acknowledge support from the Army Research Office (grant no. W911NF-17-1-0306) and the Air Force Office of Scientific Research (grant no. MURI FA9550-21-1-0058)

\newpage
\bibliographystyle{abbrvnat}
\bibliography{library,references_MIT.bib}

\end{document}